\documentclass{aa}  

\usepackage{graphicx}
\usepackage{txfonts}
\usepackage{lipsum}
\usepackage{subcaption}         
\usepackage{lscape}             
\usepackage{placeins}           
\usepackage{longtable}
\usepackage{verbatim}
\usepackage{float}
\usepackage{array,multirow,makecell}
\usepackage{xcolor, soul}
\usepackage{threeparttablex}
\usepackage[export]{adjustbox}
\usepackage{dirtytalk}
\usepackage[autostyle]{csquotes}
\usepackage[version=4]{mhchem}
\usepackage{silence}
\newcommand{\pcms}{cm$^{-2}$}
\newcommand{\pcmc}{cm$^{-3}$}
\newcommand{\kms}{km s$^{-1}$}

\defcitealias{bouvier2024}{B24}
\usepackage{natbib}
\bibpunct{(}{)}{;}{a}{}{,} 
\usepackage{hyperref}
\hypersetup{
    colorlinks=True,
    citecolor=blue,
    urlcolor=magenta,      
    linkcolor=blue,
    }

\begin{document}

   \title{Sulphur within the extreme environment of the central molecular zone of NGC\,253: a chemical modelling approach}
\titlerunning{Modelling sulphur-bearing species towards NGC\,253}
   \subtitle{}


   \author{M. Bouvier\inst{1}
   \and K. M. Dutkowska \inst{1}
        \and S. Viti\inst{1,2,3}
        \and J. G. Mangum \inst{4}
        \and E. Behrens \inst{5,6}
        \and C. Eibensteiner \inst{4}\thanks{Jansky Fellow of the National Radio Astronomy Observatory}
        }

   \institute{Leiden Observatory, Leiden University, P.O. Box 9513, 23000 RA Leiden, The Netherlands\\
             \email{bouvier@strw.leidenuniv.nl}
    \and 
   Transdisciplinary Research Area (TRA) ‘Matter’/Argelander-Institut f\"ur Astronomie, University of Bonn
   \and
    Physics and Astronomy, University College London, UK
    \and
    National Radio Astronomy Observatory, 520 Edgemont Road, Charlottesville, VA 22903-2475, USA
    \and
    Department of Astronomy, University of Virginia, P.~O.~Box 400325, 530 McCormick Road, Charlottesville, VA 22904-4325, USA
    \and
    Max-Planck-Institut für Astronomie, Königstuhl 17, D-69117, Heidelberg, Germany
    } 
   \date{}

 
  \abstract
   {
   Sulphur-bearing species are ubiquitous in Galactic star-forming regions, from dense cold cores to outflows and shocks linked to protostellar activity. A recent observational study investigated the origin of Sulphur-bearing species towards the central molecular zone (CMZ) of NGC\,253 and showed that Sulphur emission is linked to the presence of forming stars. However, more extensive modelling of these observations are required to determine the exact origin of the Sulphur emission (e.g. shock or thermal evaporation).}   
   {Using chemical modelling, we aim to further explore the origin of the emission of the sulphur-bearing species in the CMZ of NGC\,253.  We examine how Sulphur-bearing species behave in the more extreme environment of the starburst galaxy NGC 253, and more generally, how they can help us improve our understanding of the emission linked with the dense star-forming gas in external galaxies.} 
   {We use the gas-grain time-dependent chemical model \texttt{UCLCHEM} to model static warm clouds and C-type shocks under the physical conditions found in the CMZ of NGC\,253. We derive the observed abundances and abundance ratios of CS, \ce{H2S}, OCS, \ce{H2CS}, SO, \ce{SO2}, and CCS using ALMA measurements, and compare them to the modelled output abundances.}
   {We found that depending on the model type (shock, post-shock or static cloud), the highest abundance reached by the S-bearing species varies significantly. Hence, we can use the observed abundances of S-bearing species to distinguish between shocked, post-shocked or a quiescent (non-shocked) gas.
   We also confirm observationally-based conclusions on their emission origins, including in the case of unresolved emission, such as for the hot dense components of OCS and \ce{SO2}, likely probing the hot quiescent gas of proto-super star clusters within the GMCs of NGC\,253. }
   {Comparing GMC-scale observations with chemical modelling is a powerful method to investigate and constrain the origin of molecular emission towards extragalactic star-forming regions. Sulphur-bearing species are useful to distinguish between different types of ISM components (shocked/post-shocked/quiescent gas). }


   \maketitle
\nolinenumbers 

\section{Introduction}

Giant molecular clouds (GMCs), and in particular the densest regions ($n_{\mathrm{H2}}> 10^4$ cm$^{-3}$) within them, are the birthplaces of stars in galaxies \citep[e.g.][]{Wu2005, Lada2012, evans2014}. Observational studies showed that properties of GMCs strongly depend on the local galactic environment \citep[e.g.][]{hughes2013, colombo2014, sun2020, rosolowsky2021, brunetti2024}. Hence studying these regions throughout the different environments (e.g. density, metallicity, cosmic-rays, UV and X-ray radiation fields) of external galaxies is crucial to get a complete understanding of GMCs and the star formation processes, and ultimately, of the formation and evolution of galaxies \citep{SL2024}. Yet, observing star-forming regions in nearby galaxies is challenging, as multiple components of the dense interstellar medium (ISM) are mixed within a single resolution element, even at high angular resolution.

Since the chemistry in galaxies is highly influenced by the local environment, we can use observations of different molecular species and transitions arising from different environments to interpret the emission of unresolved structures. Past studies coupling radiative transfer and chemical models have been successful in interpreting the observed molecular emission from nearby galaxies \citep[e.g.][]{bisbas2014, viti_molecular_2014, kazandjian2016, scourfield_alma_2020}. Recent examples toward the central molecular zone (CMZ) of NGC\,253 using observations from the large programme ALMA Comprehensive High-resolution Extragalactic Molecular Inventory (ALCHEMI; \citealt{martin_alchemi_2021}) include CCH \citep{holdship_distribution_2021}, HCO$^+$ and HOC$^+$ \citep{harada_starburst_2021}, HCN and HNC \citep{behrens_tracing_2022}, SiO and HNCO \citep{huang_reconstructing_2023}, and HCNH$^+$ \citep{gong2025}.

ALCHEMI delivered the most complete extragalactic molecular inventory of the CMZ of this galaxy, with about 78 species detected (\citealt{martin_alchemi_2021, haasler_first_2022}). Among the detected molecules, sulphur-bearing species are highly reactive once in the gas phase. Hence, their abundances, and more generally the sulphur chemistry,  are sensitive to the thermal and kinetic properties of the gas \citep{viti_evaporation_2004}. In our Galaxy, sulphur-bearing molecules have thus been generally adopted as good tracers to investigate properties of star-forming regions, including shocks \citep[e.g.][]{bachiller_chemically_2001, codella_chemical_2005, lefloch_shock-induced_2005, podio_molecular_2014,  kwon_kinematics_2015, holdship_sulfur_2019, ospina-zamudio_molecules_2019, taquet_seeds_2020, feng_seeds_2020, tychoniec_2021, de_la_villarmois_perseus_2023, Fontani2024, Hsu2024}. In a previous study, \citealt{bouvier2024} (hereafter \citetalias{bouvier2024}) observed seven sulphur-bearing species towards ten GMCs located in the CMZ of NGC\,253: carbon monosulfide (CS), sulphur monoxide (SO), carbonyl sulfide (OCS), thioformaldehyde (\ce{H2CS}), hydrogen sulfide (\ce{H2S}), thioxoethenylidene (CCS), and sulphur dioxide (\ce{SO2}). Using the ALCHEMI observations, giving us access to several transitions covering a large range of upper energy levels, they constrained the physical conditions of the gas ($n_{\mathrm{gas}}, T_{\mathrm{gas}}$) as well as the size of the emission region for most of the sulphur-bearing species. \citetalias{bouvier2024} compared the physical conditions of the gas probed by the sulphur-bearing species with previous studies of other tracers (e.g. CCH, HCN, HNC, HNCO, SiO; \citealt{holdship_distribution_2021, behrens_tracing_2022, huang_reconstructing_2023}) and the relative abundance ratio of sulphur-bearing species with well known Galactic regions (e.g. outflows/shocks, hot cores, photodissociation regions, clouds). This led them to conclude that most of the sulphur-bearing species are likely tracing shocks throughout the CMZ. However, in several cases, they could not disentangle different potential scenarios (shock/postshock/quiescent cloud) with these methods alone. \\
\indent Past studies have combined non-LTE radiative transfer analysis with chemical modelling to interpret the molecular line emission in nearby extragalactic regions \citep[e.g.][]{Fuente2008, Rosolowsky2011, Viti2017, Harada2019, huang_reconstructing_2023, jia2026}. While in B24 a non-LTE analysis has been carried out, we now aim to complete the analysis of Sulphur-bearing species in NGC\,253 using chemical modelling.
Models of S-bearing species were first motivated by Galactic observations \citep[e.g.][]{millar_1990, charnley_sulfuretted_1997, hatchell_survey_1998, viti_chemical_2001,wakelam_resetting_2004, woods_new_2015, laas_modeling_2019}. With this work, we present new chemical models of Sulphur-bearing species in an extragalactic region, motivated mainly by 1) the need of having tailored conditions to NGC 253, (2)  the lack of shocks and (non-)thermal processes in past chemical models, and (3) a most recent and enhanced Sulphur-bearing chemical network (see Sec. \ref{sec:model}). The main goal of this work is to confirm or refute the observationally-based conclusions from \citetalias{bouvier2024} and determine the most likely origin of emission of the sulphur-bearing species. This will allow us to provide chemical templates of sulphur-bearing species in a starburst environment, as well as key information about the sensitivity of sulphur-bearing abundances under various physical conditions. We will also investigate whether sulphur-bearing species can be used as diagnostic tools to study extragalactic star-forming regions.\\
\indent The paper is organised as follows: The modelling approach is presented in Section~\ref{sec:model}, followed by the derivation of abundances and abundance ratios in Section~\ref{sec:obs}. We describe and discuss the results in Section~\ref{sec:results_discussion} before summarising our findings in Section~\ref{sec:conclusions}.

\section{Astrochemical modelling}\label{sec:model}

\begin{figure}
    \centering
    \includegraphics[width=1\linewidth]{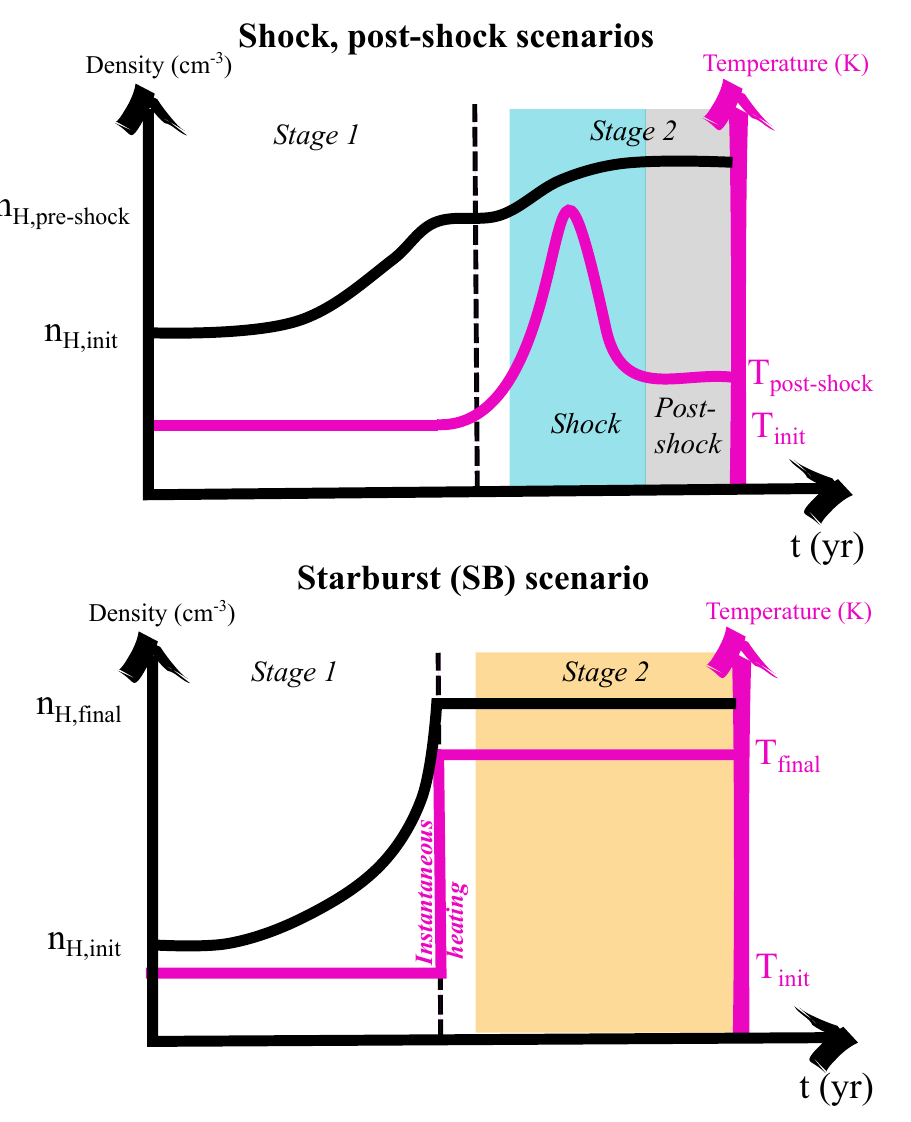}
    \caption{The shock and starburst scenarios that were modelled in this work. The density (black) and temperature (magenta; gas and dust temperatures are coupled) profiles are shown for both stages, with the transition from Stage 1 to Stage 2 indicated by a dashed black line. The blue and purple filled area represents the time range for which the models were considered in the analysis.}
    \label{fig:scheme}
\end{figure}

We used the open-source time-dependent gas-grain chemical code \texttt{UCLCHEM}\footnote{\url{https://uclchem.github.io/} version \texttt{v3.5.3}} \citep[][]{holdship2017, Vermarien2026} to model sulphur-bearing species under varying physical conditions, mimicking those in the CMZ of NGC\,253. We focussed on the sulphur-bearing species studied in \citetalias{bouvier2024}. We used the most recent version (Rate22) of the UMIST Database for Astrochemistry (UDfA; \citealt{Millar2024}) and the extended grain chemical network from \cite{Dutkowska2025} (which includes CCS, a species that is not part of the default chemical network of \texttt{UCLCHEM}.). All models include two evolutionary stages. A summary scheme of the various Stages 1 and 2 modelled is provided in Figure~\ref{fig:scheme}. 

\subsection{Stage 1 - Collapsing cloud}
Stage 1 starts from a diffuse cloud ($n_{\ce{H2}}\sim10^2$ \pcmc and $A_\mathrm{v,\ start}=2$ mag) which undergoes an isothermal collapse to a predefined final density ($n_{\ce{H},\ \mathrm{final}}$). 
The initial molecular abundances are set to zero, except for atomic elements. The elemental abundances are those from \cite{asplund_2009, jenkins2009}, except for Si, which is depleted by a factor of 100 in the starburst scenario, following \cite{Dutkowska2025}. For sulphur, we ran two sets of models, one with the elemental abundance from \cite{jenkins2009} (which is already slightly depleted to account for the fact that sulphur is observed to be depleted onto dust grains in the ISM) and one with a depletion of a factor 10, to account for a possible larger depletion, as found in Galactic star-forming regions (e.g. \citealt{bachiller_shock_1997, anderson_new_2013, holdship_h2s_2016, holdship_sulfur_2019}).

For Stage 1, we set a cosmic-ray ionization rate (CRIR) of $\zeta_0=1.31\times 10^{-17}$ s$^{-1}$ \citep[e.g.][]{caselli1998}, and a UV field (FUV) $G_0=1$ Habing\footnote{The FUV interstellar radiation field is $G_0=1.6\times10^{-3}$ erg s$^{-1}$\pcms \ \citep{habing1968}.}. It is worth noting that we also tested models with high CRIR from Stage 1, to test a scenario where the cloud collapse occurs in a highly irradiated environment. However, the final abundances from Stage 2, used in this work, were not affected. Since we do not know the initial state of the medium and to allow species to form under normal conditions, we therefore used a Galactic CRIR for Stage 1. Due to the prevailing conditions of the CMZ of NGC\,253 (high CRIR and temperature), we assume that the initial dust and gas temperature are at least 15~K. We also test initial temperatures of 20~and 25~K to examine how the Sulphur chemistry is affected by changes in initial temperatures. The need for higher initial temperature is also supported by observations showing that gas temperatures in the CMZ are high ($\geq 85$~K; \citealt{tanaka_2023}).The evolution of the gas temperature and density throughout the Stages 1 and 2 are shown for the two types of models in Figure~\ref{fig:scheme}. For the shock scenario, we let the chemistry evolve in Stage 1 up to $t=6.5\times10^6$ yr before Stage 2 starts, following \cite{Dutkowska2025}. In the starburst scenario, Stage 1 ends once the final density is reached. In both cases, the final chemical composition at the end of Stage 1 is then used as an input for Stage 2.

\subsection{Stage 2 - C-shocks and static clouds}
In Stage 2, we modelled both C-type shocks (Fig.~\ref{fig:scheme}, top plot) and starburst (SB; Fig.~\ref{fig:scheme}, bottom plot) scenarios based on our hypotheses concerning the region of emission of sulphur-bearing species (see \citetalias{bouvier2024}), and with varying conditions (see Table~\ref{tab:params} for a summary). For both types of Stage 2, we varied the CRIR with values of $\zeta/\zeta_0=10^3$ and $10^4$, representative of the values found in the CMZ of NGC\,253 (e.g. \citealt{behrens_tracing_2022, Behrens2024}). the CRIR increases instantaneously from Stage 1 to Stage 2. Stage 2 is run for $10^6$ yr. The different modelled scenarios are represented by cyan (shocks), grey (post-shocks), and orange (SB) areas in Fig.~\ref{fig:scheme}: For the shock scenario, we consider models with $T_{\mathrm{gas, dust}}>T_{\mathrm{init}}$, which corresponds to the passage of the shock, and $T_{\mathrm{gas, dust}}\neq 50$ K, which corresponds to the post-shocked scenario ($T_{\mathrm{gas,dust}}=T_{\mathrm{post-shock}}$=50 K). For the SB scenario, we consider models with $t\geq 10^4$ yr, where the chemical equilibrium is reached. 

\subsubsection{C-shock scenario}
For the shock models, we explored low to moderate shock velocities ($v_\mathrm{shock})$, between 5 and 45 \kms propagating in a strong magnetic field (see \citealt{Dutkowska2025} for details). We do not consider larger shock velocities as \citetalias{bouvier2024} found that sulphur-bearing species were likely not tracing fast shocks. We explored pre-shock densities ($n_{\ce{H},\ \mathrm{pre-shock}}$) between $10^4-10^6$ \pcmc, which covers the range of densities derived in the GMCs of the CMZ of NGC\,253 \citepalias{bouvier2024}. After the shock has passed, we consider that the gas does not cool back down to the initial temperature but remains at a temperature ($T_{\mathrm{post-shock}}$) of 50 K. We thus define the post-shocked phase as the time during which the temperature has cooled down to 50 K after the shock has passed and when the chemistry of the gas is not affected by the shock any longer. The FUV is set to $G_0=100$ Habing\footnote{Increasing further the FUV is not relevant due to the high $A_\mathrm{v}$ of Stage 2 (Minimum $A_\mathrm{v}$ for a shock model with initial density of $10^4$\pcmc is 11~mag.)}. 

\subsubsection{Starburst scenario}
For the SB (thermal) process, we modelled a warm static cloud as a Stage 2, with a set of varying initial densities ($n_{\ce{H}}=10^5-10^8$ \pcmc). The static cloud has a constant density and temperature, and does not undergo an important perturbation such as high turbulence or shocks. We thus refer to this type of model to a \say{quiescent} gas (i.e. non-shocked gas) in the rest of the manuscript. The transition from Stage 1 to Stage 2 is instantaneous in terms of temperature variations. This approach is suitable in the case of extragalactic studies, since the temporal and spatial effects are "diluted" on the size scales sampled by our measurements (e.g. \citealt{viti_evaporation_2004}), and allows for more efficient chemical model execution. The chemistry evolves until the equilibrium is reached (timescale of $t>10^4$ yr). We varied the temperature of Stage 2 ($T_{\mathrm{final \ (gas, dust)}}$; hereafter $T_\mathrm{final}$) between 50 and 300 K. Both the range of gas densities and gas temperatures correspond to the range of physical conditions derived observationally for the sulphur-bearing species \citepalias[see][]{bouvier2024}, which are summarised in Table~\ref{tab:groups}. We note that for some of the species (e.g. the low-J lines of CS, \ce{H2CS}, \ce{OCS}, SO and CCS), the densities derived from observations were lower than $10^5$ \pcmc in some of the GMCs. To model these particular species, we ran some additional ``low-density'' models with $n_H=10^3-10^4$ \pcmc, which are presented in Appendix~\ref{appdx:low_dens_models}. The FUV is set to $G_0=1000$ Habing, to model a high-star formation environment, where an enhanced UV field is expected \citep[e.g.][]{Schneider2020}. 

\begin{table}
    \caption{Parameter space covered in the models.}
    \label{tab:params}
    \centering
    \begin{tabular}{l|l}
    \hline \hline
    Parameter (Unit) & Values \\
    \hline
    \multicolumn{2}{c}{Stage 1: Collapse}\\
    \hline
    $n_{\ce{H},\ \mathrm{init}}$\tablefootmark{a} (\pcmc) & $10^2$ \\
    $n_{\ce{H},\ \mathrm{final}}$\tablefootmark{a} (\pcmc) & $10^4, 10^5, 10^6, 10^7, 10^8$ \\
    $T_{\mathrm{init}}$ (K) & 15, 20, 25 \\
    $\zeta/\zeta_0$ & 1 \\
    $G_0$ (Habing) & 1 \\
    $A_\mathrm{v,start}$ (mag) & 2\\
    $X(\ce{Si})$\tablefootmark{b} & $1.78\times10^{-8}$, $1.78\times10^{-6}$\\
    $X(\ce{S})$\tablefootmark{c} & $3.51\times10^{-7}$, $3.51\times10^{-6}$ \\
    $R_{\mathrm{final}}$\tablefootmark{d} (pc) & 0.5 \\ 
    \hline
    \multicolumn{2}{c}{Stage 2: C-Shocks}\\
    \hline
    $n_{\ce{H},\ \mathrm{pre-shock}}$\tablefootmark{a} (\pcmc) & $10^4, 10^5, 10^6$ \\
    $v_\mathrm{shock}$ (\kms) & $5, 15, 30, 45 $\\
    $T_{\mathrm{post-shock}}$ (K) & 50 \\
    $G_0$ (Habing) & 100 \\
    $\zeta/\zeta_0$ & $10^3, 10^4$ \\
    \hline
    \multicolumn{2}{c}{Stage 2: Static Cloud}\\
    \hline
    $n_{\ce{H}}$\tablefootmark{a} (\pcmc) & $10^3$\tablefootmark{e}, $10^4$\tablefootmark{e}, $10^5, 10^6, 10^7, 10^8$ \\
    $T_{\mathrm{final \ (gas, dust)}}$ (K) & $50, 100, 150, 200, 250, 300 $\\
    $G_0$ (Habing) & 1000 \\
    $\zeta/\zeta_0$ & $10^3, 10^4$ \\
    \hline
    \end{tabular}
    \tablefoot{
    \tablefoottext{a}{$n_{\ce{H}}$ is the proton density, defined as $n(\ce{H)}+2n(\ce{H2)}$;}
    \tablefoottext{b}{The SB models use collapse models with depleted \ce{Si} abundance (e.g. Savage \& Sembach 1996), while the shock models use those with the standard value of $1.78\times10^{-6}$ (Jenkins 2009);}
    \tablefoottext{c}{We considered both collapse models with depleted (ISM value of $3.51 \times 10^{-6}$; \citealt{jenkins2009}) and heavily depleted (by a factor 10) sulphur (see text).}
    \tablefoottext{d}{A single radius is adopted across all models.}
    \tablefoottext{e}{Only for the ``low-density'' models. }
    }
\end{table}


\section{Abundances and abundance ratios}\label{sec:obs}
A key result from \citetalias{bouvier2024} was that for most sulphur-bearing species located in the inner GMCs (GMC 3-7), two gas components (comp1 and comp2) were needed to fit the data: the low energy transitions did not trace the same ISM component as the higher energy transitions. This is the case for all species except \ce{H2S}, for which only three transitions were available. The case of SO is particularly interesting because the gas conditions traced by the low-E$_{\rm{u}}$ lines drastically change between the outer and inner GMCs. We thus separate the gas traced by the low-E$_{\rm{u}}$ lines in the outer GMCs (SO$_{\mathrm{comp1a}}$) from that in the inner GMCs (SO$_{\mathrm{comp1b}}$). For each species and their component, except in the cases of CCS, \ce{SO2}, and the high-E$_{\rm{u}}$ component of OCS, \citetalias{bouvier2024} ran non-LTE analyses which allowed them to derive the size of the emitting region (linked with the beam filling factor; see Tables 3 and E.1 from \citetalias{bouvier2024}). Overall, three different size regions were derived: $\theta \geq 1.6''$ ($\sim$ 27 pc; corresponding to $\sim$ GMC scales), $\theta \sim 0.6''$ ($\sim 10$ pc), and $\theta \sim 0.2''$ ($\sim 3$ pc, corresponding to groups of proto-super star clusters or pSSCs, identified in the inner GMCs 3 to 6; e.g. \citealt{ando_2017, leroy_forming_2018,rico-villas_super_2020, mills_clustered_2021, levy_outflows_2021}). A summary of each species, components and their region of emission is shown in Table~\ref{tab:groups}.

\begin{table*}[]
    \centering    
    \caption{Species and components, GMCs where they are detected, physical conditions ($T_\mathrm{gas}, n_\mathrm{gas}$) derived in \citetalias{bouvier2024}, and derived abundance range. Favoured models from abundance comparison (see Sec.~\ref{subsec:tools}) are also indicated. }
    \label{tab:groups}
    \resizebox{\linewidth}{!}{
    \begin{tabular}{lcccccc}
    \hline\hline
        Species/component & GMCs &$T_\mathrm{gas}$ range & $n_\mathrm{gas}$ range& Region of emission   & Abundance range & Favoured model? \\
        & & (K) & (\pcmc)&($''$) & &  SB/S/PS/None\tablefootmark{d}\\
        \hline
        CS$_{\mathrm{comp1}}$& All &5 -- 14& $\geq 3\times10^3$& $\geq 1.6$ &$2\times10^{-10}-7.8\times10^{-8}$& None \\
        CS$_{\mathrm{comp2}}$& All &15 -- 110 &$\geq 1.5\times10^5$&$0.6$ &$7\times10^{-12}-5.6\times10^{-9}$ & SB/PS\\
        SO$_{\mathrm{comp1a}}$& outer GMCs\tablefootmark{a} &30 -- 300&$(0.5-10)\times10^4$& $0.2$ &$1\times10^{-12}-2.4\times10^{-8}$& None\\
        SO$_{\mathrm{comp1b}}$& inner GMCs\tablefootmark{b} &11 -- 300& $\geq 8\times10^4$& $0.6$ & $7\times10^{-12}-9.5\times10^{-10}$ & SB\\
        SO$_{\mathrm{comp2}}$& inner GMCs &27 -- 260& $\geq 3\times10^5$& $0.2$ & $9\times10^{-13}-2.4\times10^{-9}$& SB/PS\\
         \ce{H2S} &  GMC 10 + inner GMCs &30 -- 159&$\geq 2\times10^6$& 0.2 & $2\times10^{-11}-7.1\times10^{-9}$ &S\\
         OCS$_{\mathrm{comp1}}$ & All &12 -- 100 &$\geq 2\times10^4$&0.2  &$1\times10^{-11}-2.4\times10^{-7}$& None \\
         OCS$_{\mathrm{comp2}}$ & All &124 -- 312&...& 0.2\tablefootmark{c} & $4\times10^{-11}-1.4\times10^{-9}$ & SB/PS\\
         \ce{H2CS}$_{\mathrm{comp1}}$ & All &10 -- 70&$\geq 10^3$& 0.2 &$4\times10^{-13}-9.5\times10^{-8}$ &None\\
        \ce{H2CS}$_{\mathrm{comp2}}$ & All &50 -- 300&$(0.1 -10)\times10^6$& 0.2 & $2\times10^{-12}-1.7\times10^{-8}$ &None\\
         CCS & All but GMC 10 &5 -- 72 &...&1.6/0.6/0.2\tablefootmark{c} &  $4\times10^{-12}-4.8\times10^{-10}$& S\\
         \ce{SO2} & inner GMCs  &55 -- 298 &...&0.2\tablefootmark{c} & $2\times10^{-11}-9.5\times10^{-10}$& SB/PS\\
         \hline
    \end{tabular}}
    \tablefoot{
    \tablefoottext{a}{Outer GMCs are GMC 1a,1b, 2b, 8a, 9a, 10}
    \tablefoottext{b}{Inner GMCs are GMC 3, 4, 6, 7}
    \tablefoottext{c}{Value assumed not derived (see text)}
    \tablefoottext{d}{SB, S, and PS refer to starburst, shock and post-shock models, respectively.}
    }
\end{table*}

In the case of CCS, \ce{SO2}, and OCS$_{\mathrm{comp2}}$, \citetalias{bouvier2024} could not derive the size of the region of emission from the LTE analysis. For CCS, since it can be detected in a large range of environments from cold dark clouds\citep[e.g.][]{hirota_2009, vastel_sulphur_2018, giani2025} to shocks \citep[e.g.][]{holdship_sulfur_2019, nakamura2024} and high-mass protostars \citep[e.g.][]{tercero_line_2010, fontani_evolution_2023, chen2025},
we assume it comes from any of the three regions of emission. This is supported by the fact that emission of this species is resolved at low-E$_{\rm{u}}$ but unresolved at high-E$_{\rm{u}}$ \citepalias{bouvier2024}. For OCS$_{\mathrm{comp2}}$, since the emission distribution of the high-E$_{\rm{u}}$ lines is more compact compared to the low-E$_{\rm{u}}$ lines, we assume that it arises mostly from the most compact regions, at the scales of pSSCs (0.2 pc). Finally, in the case of SO$_2$, although \citetalias{bouvier2024} found that two excitation temperatures are needed (hinting at two different ISM components), the emission distribution of the various transitions is always compact towards the inner GMCs (see Figures 2 and 3 from \citetalias{bouvier2024}). We thus assume that the SO$_2$ emission arises mostly from the smallest scales (0.2 pc). 

We first calculate the observed fractional abundances (i.e. with respect to \ce{H2}). We consider that the physical process at the origin of the emission of each sulphur-bearing species is the same in all the GMCs. Hence, we use here a unique range of observed abundances for all the 10 GMCs. Derived abundances are shown in Table~\ref{tab:groups}, and a detailed description of the abundance calculations are presented in Appdx.~\ref{subsec:abun}.

\subsection{Ratios}\label{subsec:ratios}

Following the results in \citetalias{bouvier2024} we only consider abundance ratios between sulphur-bearing species and components that belong to the same region of emission. For simplicity, we assume that for each species, one component traces the same type of gas throughout the CMZ. We note that there could be variations from one GMC to another, but investigating the differences between GMCs is out of the scope of this paper. Differences between the outer and inner GMCs are accounted for with the presence of a second component. 
We selected abundance ratios for each region of emission based on the hypotheses derived from the observations \citepalias[][]{bouvier2024}. 
For species emitting on the largest scales ($\sim 1.6''$), while \citetalias{bouvier2024} concluded that CS$_{\mathrm{comp1}}$ traces quiescent dense gas, they could not determine whether CCS probes quiescent dense gas or shocks.
For species emitting on intermediate scales ($\sim 0.6''$), \citetalias{bouvier2024} concluded that SO$_{\mathrm{comp1a,b}}$ likely traces dense compact gas while for CCS and CS$_{\mathrm{comp2}}$ both a quiescent dense gas and a shock origin could fit the data. 
For species emitting on the smallest scales ($\sim 0.2''$), \ce{H2S}, OCS$_{\mathrm{comp1}}$, \ce{H2CS}$_{\mathrm{comp1}}$ and SO$_{\mathrm{comp2}}$ were proposed as potential shock tracers \citepalias{bouvier2024}. For \ce{H2CS}$_{\mathrm{comp2}}$, SO$_2$, and OCS$_{\mathrm{comp2}}$, both a shock or a hot gas (due to thermal evaporation from the grain ice mantles) origin fit the data. For CCS and SO$_{\mathrm{comp1a}}$, the hypothesis remains similar: SO probes dense quiescent gas and CCS probes dense quiescent gas or shocks. The abundance ratios and their related hypotheses investigated are presented in Table~\ref{tab:ratios}.

\begin{table*}[]
    \centering
        \caption{Range of abundance ratios derived for each region of emission ($\theta$) and for the GMCs and tested hypotheses.}
    \label{tab:ratios}
    \resizebox{\linewidth}{!}{
    \begin{tabular}{clclc}
    \hline \hline
     $\theta ('')$ & Ratio   & Measured range   &Possible origins\tablefootmark{a} & Models tested (SB/S/PS)\tablefootmark{b}  \\
      \hline
      1.6 &CS$_{\mathrm{comp1}}$/CCS   & 0.7 -- $2.9\times10^3$   &Both species trace quiescent dense gas or different environments & All\\
         \hline
        \multirow{3}{*}{0.6} & CS$_{\mathrm{comp2}}$/CCS & 0.04 -- 224.0 & Both species trace quiescent/shocked/post-shocked gas & All \\
        & CS$_{\mathrm{comp2}}$/SO$_{\mathrm{comp1b}}$ & 0.02--70.0& Both species trace dense quiescent gas & SB  \\
        \hline
         \multirow{10}{*}{0.2}   & OCS$_{\mathrm{comp1}}$/\ce{H2CS}$_{\mathrm{comp1}}$&$4\times10^{-3}-2\times10^4$& Both species trace shocked/post-shocked gas & S/PS \\
         &\ce{H2S}/OCS$_{\mathrm{comp1}}$ & $2\times10^{-3}-94.3$ & Both species trace shocked/post-shocked gas & S/PS\\
         & \ce{H2S}/SO$_{\mathrm{comp2}}$&0.1 -- $4.2\times10^3$  &Both species trace shocked/post-shocked gas& S/PS \\
         & OCS$_{\mathrm{comp1}}$/CCS &0.2 -- 884  & Both species trace shocked/post-shocked gas& S \\
         & SO$_{\mathrm{comp2}}$/CCS&$2\times10^{-3}-393$ & Both species trace shocked/post-shocked gas& PS\\
         &\ce{OCS}$_{\mathrm{comp1}}$/SO$_{\mathrm{comp1a}}$&$4\times10^{-3}$ -- $10^4$& SO$_{\mathrm{comp1a}}$ does not trace shocked/post-shocked gas & SB$^*$\tablefootmark{c}\\
         &\ce{H2CS}$_{\mathrm{comp1}}$/SO$_{\mathrm{comp1a}}$&$2\times10^{-4}-1.3\times10^3$& Both species trace post-shocked gas & PS\\
         & OCS$_{\mathrm{comp2}}$/\ce{SO2}& $0.06-26$& Both species trace shocked or warm quiescent gas& All\\
         & OCS$_{\mathrm{comp2}}$/\ce{H2CS}$_{\mathrm{comp2}}$& $6\times10^{-3}-299$&Both species trace  shocked or warm quiescent gas & All\\
         &\ce{H2CS}$_{\mathrm{comp2}}$/\ce{SO}$_{\mathrm{comp2}}$ & $7\times10^{-3}-310$&Both species probe  warm quiescent gas & SB/PS\\
         & \ce{H2S}/OCS$_{\mathrm{comp2}}$ &0.3 -- 196 & Both species trace shocked gas & S\\
         &\ce{H2S}/\ce{SO2} &0.4 -- 295 & Both species trace shocked gas & S\\
         \hline
    \end{tabular}}
    \tablefoottext{a}{based on \citetalias{bouvier2024}.}
    \tablefoottext{b}{SB, S, and PS refer to starburst, shock and post-shock models, respectively.}
    \tablefoottext{c}{Low density models only.}
\end{table*}

\section{Results and discussion}\label{sec:results_discussion}

The purpose of this work is to determine the most likely origin of emission of sulphur-bearing species towards the CMZ of NGC\,253. Hence, when comparing models with observations, we focus on the most meaningful agreement from a statistical perspective. Therefore, for the violin plots (Figure \ref{fig:comp_models_obs}), we consider the models to be in good agreement with the observations if at least the median (50th percentile of the data) falls within the observation range. The agreement is better when the full interquartile range (representing the middle 50 percent of the models) aligns with the observations. Since here the width of the violin plot is proportional to the number of models, we also consider a good agreement with the observations if the widest part of the violin plot falls within the observational range. In contrast, if only a part of the interquartile range (without the median), a narrow tail outside the interquartile range or a narrow area of the violin plot is consistent with the observations, we do not consider it as a good agreement. For the letter-value (boxen) plots \citep[Figures \ref{fig:groupA} through \ref{fig:Ratio_groupC_SB},][]{letter-value-plot}, the median and quantiles ranges are indicated, with the width associated to the corresponding percentile: The first box around the median represents 50\% of the models, the next box around the median contains 50\% of the remaining data (hence 25\%) and so on. Hence, we consider that there is a very good agreement with the observations when the first box (containing 50\% of the models) falls within the observational range. If this is not the case, no good agreement is found.

\subsection{Sulphur-bearing species as diagnostic tools}\label{subsec:tools}
\begin{figure*}
    \centering
    \includegraphics[width=0.9\linewidth]{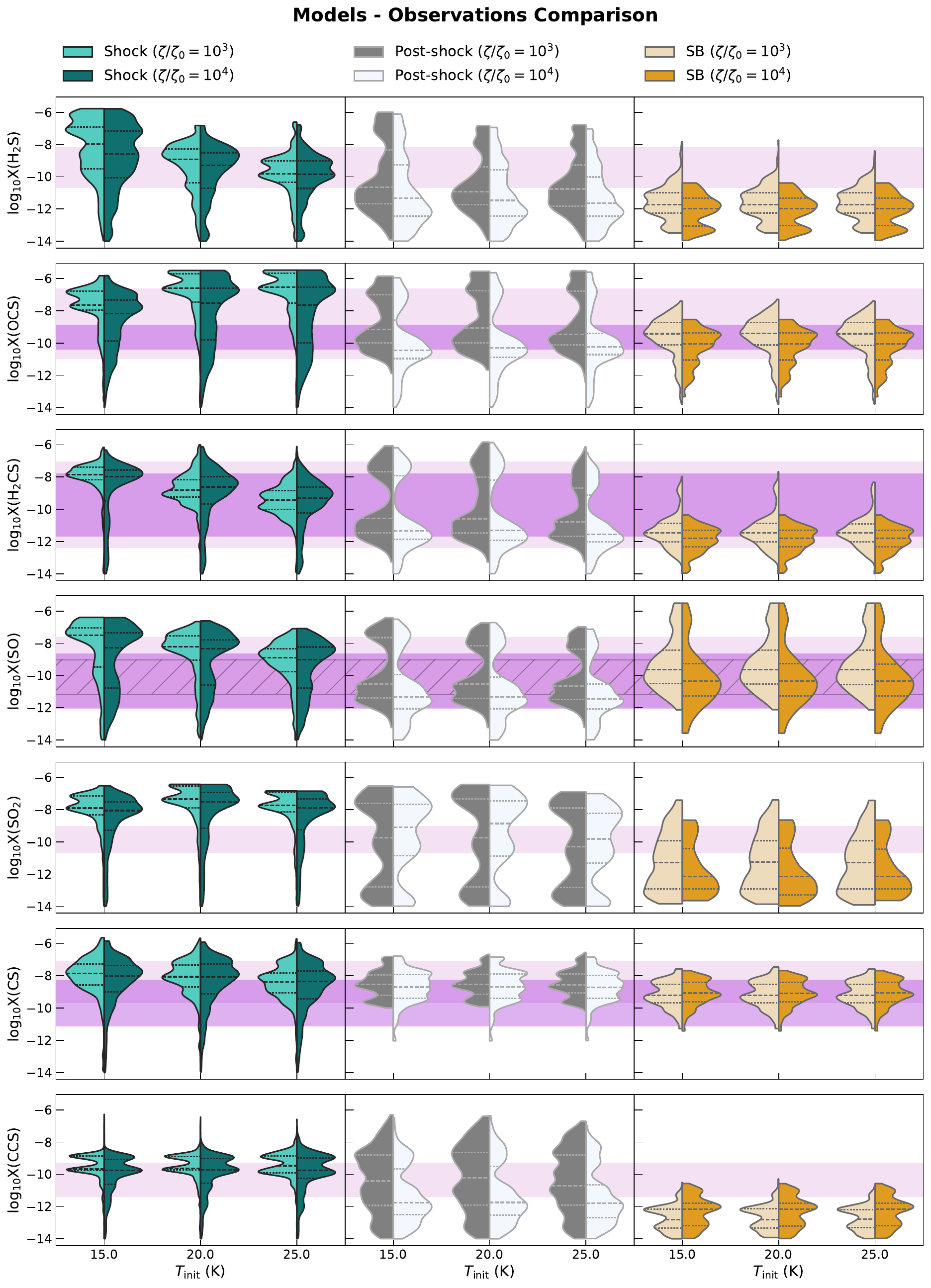}
    \caption{Violin plots of the fractional abundances in shock (cyans), post-shock (greys) SB (oranges) models as a function of the initial temperature (x-axis) and $\zeta/\zeta_0$ (shades of colours). The interquartiles of the data (representing the middle 50\% of the distribution) are shown. The observed abundances are shown in a horizontal shaded purple area. For species presenting a second component (comp2), the shaded area is darker. In the SO case (see text), we distinguish SO$_{\mathrm{comp1a}}$ (light purple) from SO$_{\mathrm{comp1b}}$ (dashed area).}
    \label{fig:comp_models_obs}
\end{figure*}

In this section we investigate (1) which sulphur-bearing species preferentially trace different dense gas environments and (2) which sulphur-bearing species can be used as environmental diagnostic tools, and under which conditions. We compare the modelled fractional abundances of each species in different models: starburst (SB), shock and post-shock, as shown in Figure~\ref{fig:comp_models_obs}. Comparing the various models with the observations, we see that:\\

$-$ For \ce{H2S}, shock models are in excellent agreement with the observations, while for CCS, both shock models and post-shock models are in good agreement with the observations. On the other hand, the majority of the SB models generally under-produce \ce{H2S} and CCS.\\
$-$ Shock models are in better agreement with the observations for \ce{H2CS}$_{\mathrm{comp2}}$, compared to post-shock and SB models.\\
$-$ For \ce{OCS}$_{\mathrm{comp2}}$, \ce{SO}$_{\mathrm{comp1b}}$, \ce{SO}$_{\mathrm{comp2}}$, \ce{SO2}, and \ce{CS}$_{\mathrm{comp2}}$, the post-shock and SB models are favoured over the shock models.\\
$-$ The SB models agree better with the observed values compared to the post-shock models for \ce{SO}$_{\mathrm{comp1b}}$ only.\\
$-$ For \ce{OCS}$_{\mathrm{comp1}}$, \ce{H2CS}$_{\mathrm{comp1}}$ and \ce{SO}$_{\mathrm{comp1a}}$,  the range of observed abundances is too large to favour one of the models.\\

The favoured model(s) for each species and component is indicated Table~\ref{tab:groups}. In summary, no single physical model reproduces all sulphur-bearing species simultaneously: Different species favour different regimes.

Nevertheless, if the abundances are sufficiently constrained, under the extreme conditions tested in this work, most sulphur-bearing species can be useful tools to distinguish between the various types of environments (shock, post-shock and quiescent).
Shock models usually produce a higher amount of sulphur-bearing species compared to SB models (with abundances $\geq 10^{-8}$ for OCS, SO, \ce{SO2} and $\geq 10^{-10}$ for \ce{H2S}, \ce{H2CS}, and CCS). CS is the only species where the distinction between shock and SB models is not as clear as for the other sulphur-bearing species, unless derived CS abundances are above $X>10^{-7}$. 
On the other hand, the distinction between post-shock and SB models is not as clear, due to a larger spread of the range of abundances in the post-shock models, and the fact that the median of the abundances is the same in both types of models. We can distinguish between the two types of models only using \ce{H2S}, OCS, \ce{H2CS} and CCS if their abundance is larger than $X\gtrsim 10^{-8}$. This is also the case for \ce{SO2} but only for the highest CRIR, $\zeta/\zeta_0=10^4$. 
Distinguishing shock and post-shocks models based on the fractional abundances of sulphur-bearing species alone is not possible: post-shock models can produce similarly high amounts of sulphur-bearing species compared to the shock models, consistent with what was found in \cite{Dutkowska2025}. This indicates that warm chemistry is still occurring in our defined post-shocked phase (which has a constant temperature of 50 K), and that freeze out is not regulating the chemistry in this phase.  Post-shock models can produce much lower amounts of \ce{H2S}, \ce{H2CS}, \ce{SO2}, and CCS. Hence, a distinction might only be possible if $X\leq 10^{-11}-10^{-12}$. Table~\ref{tab:diagnostic} shows a summary of the most favoured and disfavoured environment for the various sulphur-bearing species, based on their abundances.


Finally, it is worth noting that the initial temperature ($T_{\mathrm{init}}$) and the CRIR ($\zeta/\zeta_0$) are not critical parameters when comparing models with our observations\footnote{The dependence on the other tested parameters is presented in Appendix~\ref{appdx:params_dependence} and investigated in the next Section (Sec.~\ref{subsec:origin_emission})}. A change in the initial temperature only affect the maximum abundance reached for some species (\ce{H2S}, OCS, SO, \ce{SO2}) in shocked and post-shocked models while CRIR has an impact on the abundances of sulphur-bearing species in all SB models: The maximum abundances of \ce{H2S}, OCS and \ce{SO2} decrease with increasing CRIR, and the median abundances decrease for all the sulphur-bearing species except for CCS, where it increases with increasing CRIR. In shock and post-shock models, while the maximum abundances of the species are not impacted by the CRIR, the median abundances vary depending on the species. CS is the least impacted species by a change in CRIR and initial temperature and can thus be used as a reference standard when paired with other species sensitive to CRIR, as found by \cite{Dutkowska2025}.


\begin{table}[]
\centering
\caption{Most favoured or disfavoured environment probed by the sulphur-bearing species based on their abundance.}
\label{tab:diagnostic}
\resizebox{\linewidth}{!}{
    \begin{tabular}{ccc}
    \hline \hline
        Sulphur-bearing & favoured/disfavoured & Conditions  \\
        species & environment & \\
        \hline
        OCS, SO, \ce{SO2} & shock/quiescent & $X\geq 10^{-8}$\\
        \ce{H2S}, \ce{H2CS}, \ce{CCS} & shock/quiescent & $X\geq 10^{-10}$\\
        CS & shock/quiescent & $X>10^{-7}$\\
        \ce{H2S}, OCS, \ce{H2CS}, CCS & post-shock/quiescent & $X\geq 10^{-8}$\\
        \ce{SO2} & post-shock/quiescent & $X\geq 10^{-8}$ and $\zeta/\zeta_0=10^4$\\
        \ce{H2S}, \ce{H2CS}, \ce{SO2}, CCS & post-shock/shock & $X\leq 10^{-11}-10^{-12}$\\
        \hline
    \end{tabular}}
\end{table}

\subsection{Origin of sulphur-bearing species in NGC\,253}\label{subsec:origin_emission}

To help constrain the likely origin of emission of the various sulphur-bearing species, we investigate the abundance ratios defined in Sec.~\ref{subsec:ratios} to confirm or refute the various hypotheses drawn from the observations in \citetalias{bouvier2024}. The models used to test the various hypotheses are indicated in Table~\ref{tab:ratios}. 

\begin{figure}
    \centering
    \includegraphics[width=\linewidth]{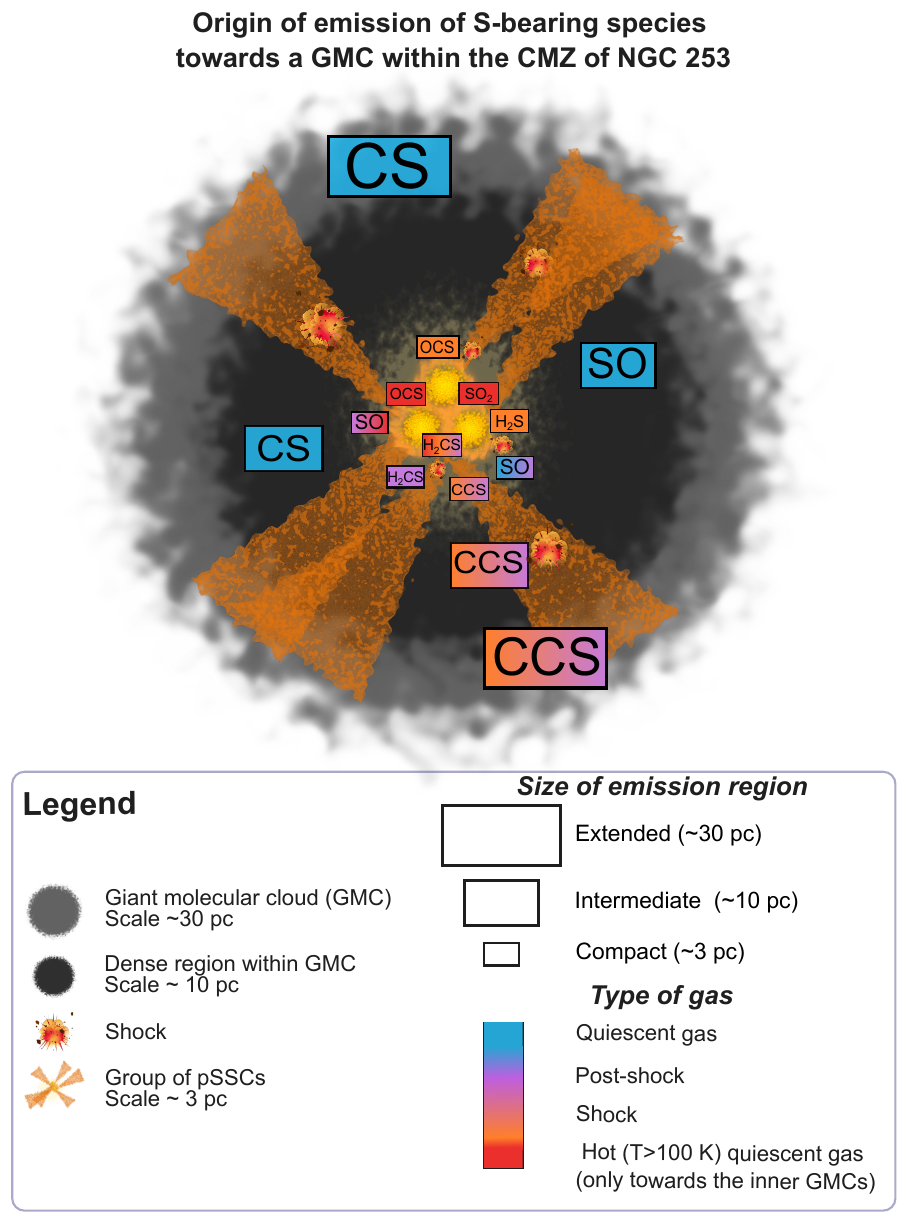}
    \caption{Scheme (not to scale) summarising the origin of emission of sulphur-bearing species towards the CMZ of NGC\,253 (see Sec.~\ref{subsec:origin_emission}), combining results from both observations \citepalias[from][]{bouvier2024} and chemical models (this work). A typical inner GMC (which includes the presence of pSSCs) is represented. In outer GMCs, hot emission of \ce{OCS}, \ce{H2CS} and \ce{SO2} are not detected. }
    \label{fig:summary_scheme}
\end{figure}

Figure~\ref{fig:summary_scheme} presents a summary schematic of our findings, updated from the one presented in \citetalias{bouvier2024}. The schematic is also simplified as it focusses only on the type of gas probed by the sulphur-bearing species towards a typical GMC of NGC\,253's CMZ. 
We discuss in details the most likely origin for each species and component in the sections below.

\subsubsection{Origin of extended species ($\sim 30$ pc scales)}

\begin{figure}
    \centering
    \includegraphics[width=1\linewidth]{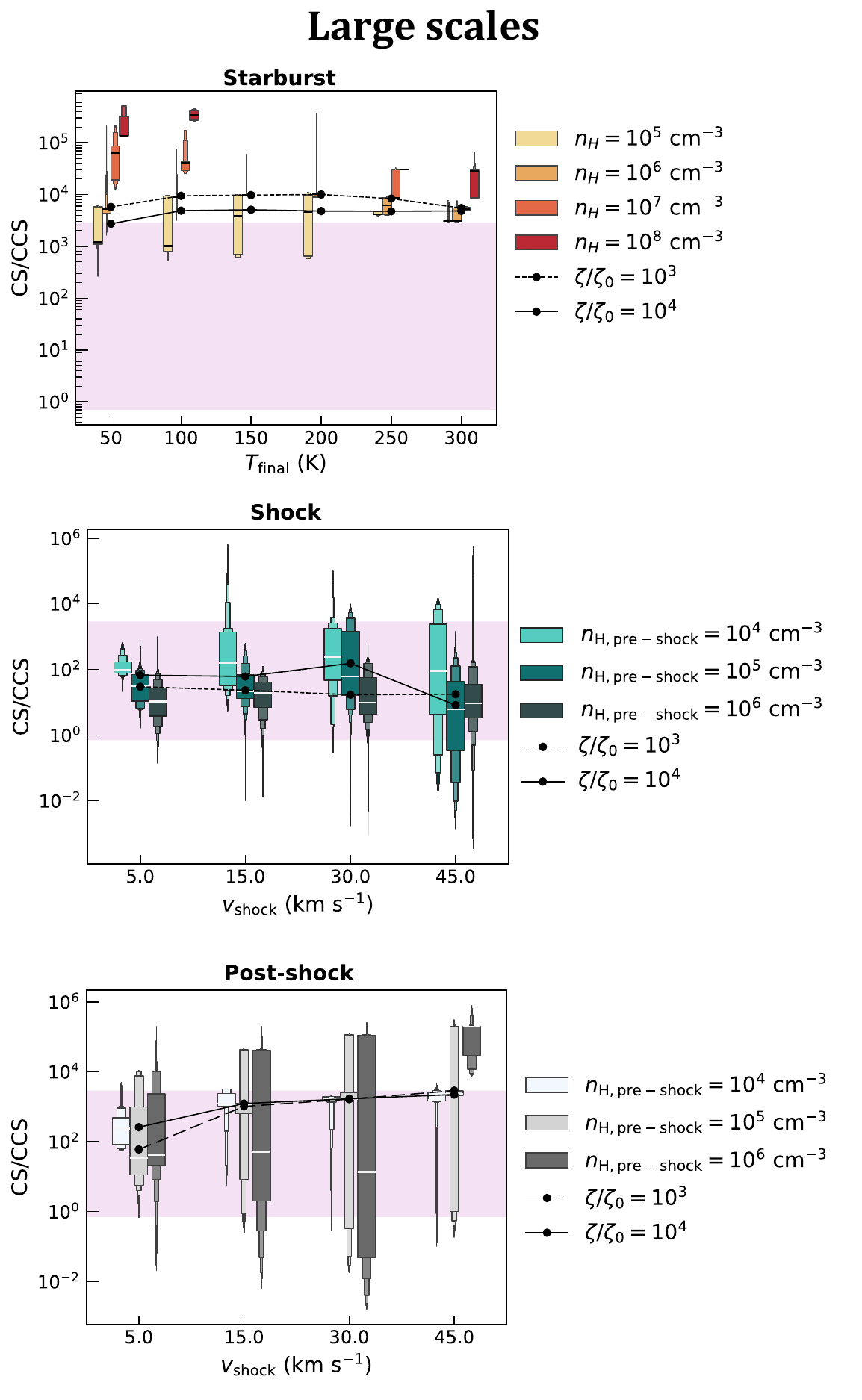}
    \caption{Boxen plots of the abundance ratio of species emitting on large (1.6$''$) scales for the SB models (oranges) as a function of the final gas temperature (x-axis) and final density (shades of colours), and for the shock (cyans) and post-shock (greys) models as a function of the shock velocity (x-axis) and pre-shock density (shaded colours). The width of the boxes corresponds to each percentile with the median (Q2, 50th percentile) highlighted by the white horizontal line. The extent of the box plots covers the full range of data. For all models, the median ratio as a function of $\zeta/\zeta_0$ is indicated by the dashed ($\zeta/\zeta_0=10^3$) and solid ($\zeta/\zeta_0=10^4$) lines. The ratio range derived from the observations is shown with the purple shaded area and corresponds to those from Table~\ref{tab:ratios}.}
    \label{fig:groupA}
\end{figure}

From the observational results, CS$_{\mathrm{comp1}}$ is likely probing dense quiescent gas while the origin of CCS was not well constrained. From Sec.~\ref{subsec:tools}, shock and post-shock models are favoured for CCS while we could not conclude for CS$_{\mathrm{comp1}}$. We thus examine whether CS$_{\mathrm{comp1}}$ and CCS could originate from the same type of environment by comparing their ratio with SB, shock, and post-shock models as shown in Figure~\ref{fig:groupA}. Since densities as low as $n_\mathrm{H2}=10^3-10^4$ \pcmc \ were found for CS$_{\mathrm{comp1}}$ \citepalias[][]{bouvier2024}, we also consider the low-density SB models in Appdx.~\ref{appdx:low_dens_models}. Overall, only the SB models with the final densities of $n_{\mathrm{H}}=10^4-10^5$ \pcmc and for a final gas temperature of $T_{\mathrm{final}}\leq 200$ K are in agreement with the observed ratio.
If we examine the individual abundances of CCS and CS in Figures~\ref{fig:low-dens-SB} and~\ref{fig:low-dens-SB-temp}, we see that they generally agree with the observations under the same conditions ($n_\mathrm{H}=10^4$ \pcmc, $\zeta/\zeta_0=10^3$) as for their ratio, but only towards the outer GMCs (lowest CRIR).
On the other hand, for shock and post-shock models, the ratio is in good agreement with the observations for all shock velocities and CRIR, except in the post-shock models where the ratio is overestimated for the highest pre-shock density ($n_{\mathrm{pre-shock}}=10^6$ \pcmc) and with $v_{\mathrm{shock}}=45$ \kms. This would imply that CS$_{\mathrm{comp1}}$ (governed by the low-J transitions of CS)  better traces shock or post-shock gas in NGC\,253. Such result would be in discrepancy with past results that showed that low-J lines of CS are a probe of dense quiescent gas \citep[e.g.][]{bayet_extragalactic_2009, aladro_2011, leroy_alma_2015}.  Hence, if we assume that the same component of a species probes the same type of gas throughout the CMZ,  CS$_{\mathrm{comp1}}$ and CCS are not probing the same gas: CS$_{\mathrm{comp1}}$ likely originates from the dense quiescent gas while CCS originates from shocked or post-shocked gas.\\
\indent We investigate further whether models match with the observations looking at the various parameters tested ($T_{\mathrm{final}}, n_{\mathrm{H}}$) in the SB models for CS$_{\mathrm{comp1}}$ and whether we can constrain the type of shock ($v_{\mathrm{shock}}, n_\mathrm{{pre-shock}}$) for CCS. From Figures~\ref{fig:SB_temp} and ~\ref{fig:SB_dens}, models with $T_{\mathrm{gas}}$ up to 250 K and $n_{\mathrm{H}}\geq 10^5$ \pcmc \ at both CRIR, statistically agree with the observed abundance of CS$_{\mathrm{comp1}}$. While the density range is in agreement with that derived in \citetalias{bouvier2024} (see also Table~\ref{tab:groups}), the temperature is much higher compared to the derived ones. This discrepancy might be related to the fact that in \citetalias{bouvier2024} the size of the emission was fixed to the size of the GMCs ($\sim 30$ pc). If CS$_{\mathrm{comp1}}$ arises from a smaller region, then the derived temperature would be underestimated. For CCS, shock models agree with the derived abundances for all pre-shock densities and shock velocities (see Figures~\ref{fig:shock_vel} and ~\ref{fig:shock_dens}), while for the post-shock models (see Figures~\ref{fig:postshock_vel} and ~\ref{fig:postshock_dens}), shock velocities of $v_\mathrm{shock}=5$ \kms \ and pre-shock densities of $n_{\mathrm{pre-shock}}=10^4$ \pcmc \ (for $\zeta/\zeta_0=10^3$, i.e. in the outer GMCs) and $n_{\mathrm{pre-shock}}=10^5$ \pcmc \ (for $\zeta/\zeta_0=10^4$. i.e. in the inner GMCs) better reproduce the observations, consistent with the fact that higher volume densities are found towards the inner GMCs \citep[e.g.][]{tanaka_2023, Behrens2024}. Therefore, if CCS traces a post-shocked gas, it is likely due to the presence of extremely slow shocks ($v_\mathrm{shock}=5$ \kms). If faster shocks are present in the CMZ of NGC\,253, then CCS is more consistent with shock emission.


\subsubsection{Origin of moderately extended species ($\sim$10 pc scales)}
\begin{figure}
    \centering
    \includegraphics[width=1\linewidth]{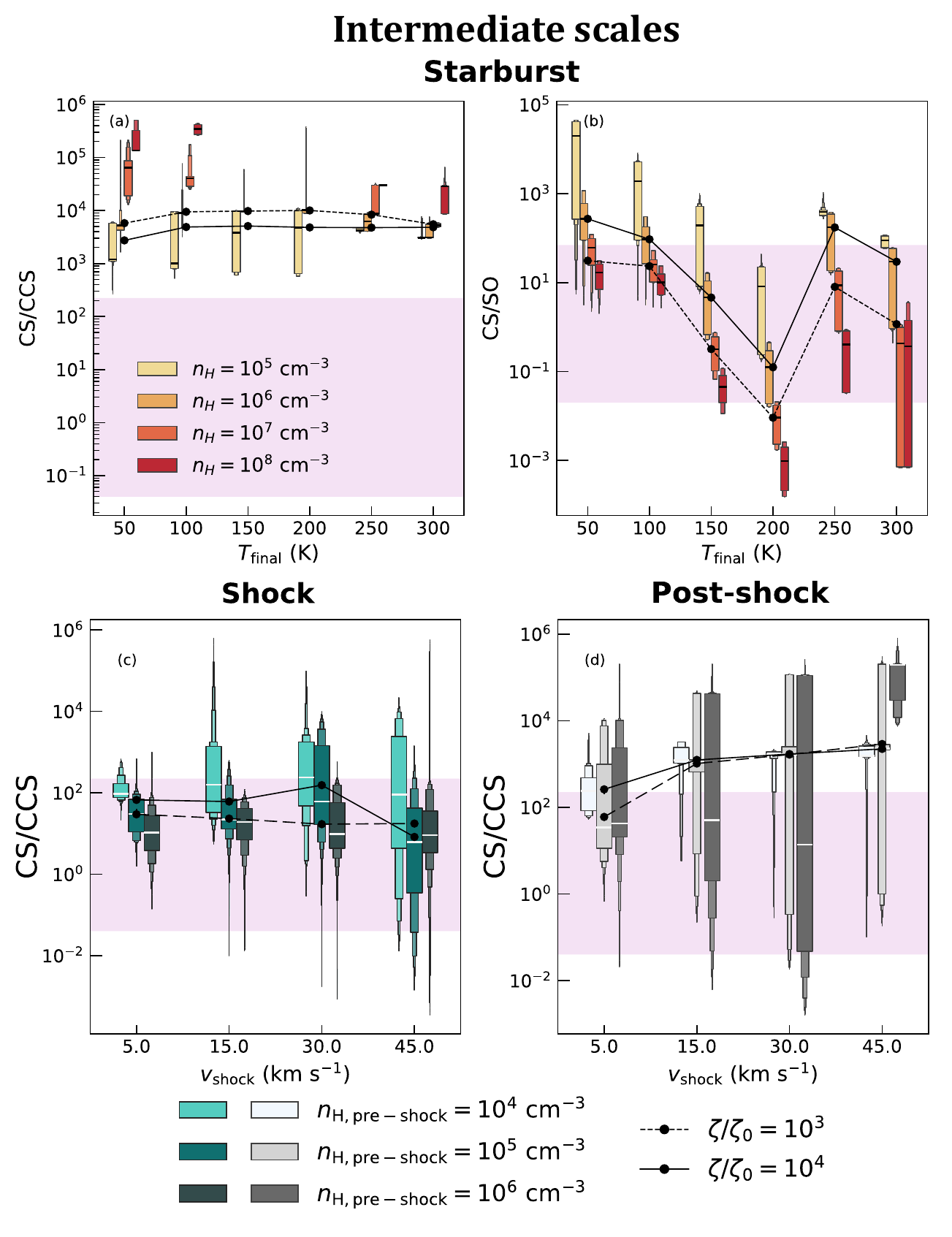}
    \caption{Same as Figure~\ref{fig:groupA} for abundance ratios of species emitting on intermediate (0.6$''$) scales.}
    \label{fig:groupB}
\end{figure}

Two ratios were defined for species emitting at intermediate scales ($\sim 10$ pc), CS$_{\mathrm{comp2}}$/CCS and CS$_{\mathrm{comp2}}$/SO$_{\mathrm{comp1b}}$. From the observations,  while SO$_{\mathrm{comp1b}}$ was found to likely probe dense quiescent gas, no conclusion was reached for CS$_{\mathrm{comp2}}$ and CCS. Hence, using the hypotheses from Table~\ref{tab:ratios}, we compare the ratio CS$_{\mathrm{comp2}}$/CCS with the observations for the three types of models in Figure~\ref{fig:groupB} (panels a, c, and d) to investigate the origin of CS$_{\mathrm{comp2}}$ and CCS. For CS$_{\mathrm{comp2}}$/SO$_{\mathrm{comp1b}}$, we only compare our observations with the SB models (panel b), to confirm the origin of SO$_{\mathrm{comp1b}}$ and investigate that of CS$_{\mathrm{comp2}}$.

We see that SB models (Figure~\ref{fig:groupB}a) always overestimate the observed CS$_{\mathrm{comp2}}$/CCS ratio. Hence, at these scales, both species clearly do not probe the same dense quiescent gas. On the other hand, the shock models (Figure~\ref{fig:groupB}c) show an excellent agreement with the observations, while in the post-shock models (Figure~\ref{fig:groupB}d), the best agreements are for $v_{\mathrm{shock}}=5$ \kms \ for all $n_{\mathrm{pre-shock}}$, or for $v_{\mathrm{shock}}=15-30$ \kms \ for $n_{\mathrm{pre-shock}}=10^6$ \pcmc. 
Figures~\ref{fig:shock_vel} to \ref{fig:postshock_dens} show how the abundances of CS and CCS behave with the various shock parameters. The modelled abundances of the two species reproduce simultaneously the observations for shocks with $v_\mathrm{shock}=5$ or $v_\mathrm{shock}=45$ \kms \ and $n_{\mathrm{pre-shock}}=10^6$ for $\zeta/\zeta_0=10^3$ or $n_{\mathrm{pre-shock}}=10^4$ for $\zeta/\zeta_0=10^4$. This would suggest that the density of the gas is higher where the cosmic-ray ionization rate is the lowest, i.e. in the external GMCs, which is opposite to what is observed. For the post-shock models, CS and CCS favour different shock velocities, with CS best reproduced if $v_\mathrm{shock}\geq 15$\kms \ and CCS at $v_\mathrm{shock}=5$ \kms. Hence, if CS$_{\mathrm{comp2}}$ traces shocked or post-shocked gas, it could be from a different shock or post-shock component from that of CCS.

However, CS$_{\mathrm{comp2}}$ probing dense quiescent gas better agrees with our result from Sec.~\ref{subsec:tools}. Comparison of modelled and observed fractional abundances also showed that the SB models statistically agree better with SO$_{\mathrm{comp1b}}$. Looking at the SB models of Fig.~\ref{fig:groupB}b, we find that depending on the final gas temperature, the best agreement between the modelled and observed CS$_{\mathrm{comp2}}$/SO$_{\mathrm{comp1b}}$ ratio does not occur at the same final densities: For $T_{\mathrm{final}}\leq 100$ K, models with $n_{\mathrm{H}}= 10^7-10^8$ \pcmc \ are in agreement with the observations. For $T_{\mathrm{final}}=150-300$ K, models with $n_{\mathrm{H}}= 10^6-10^8$ \pcmc \ are in good agreement with the observations, except at 200 K. From the observational results, common conditions where both CS$_{\mathrm{comp2}}$ and SO$_{\mathrm{comp1b}}$ emit are for $n_{\mathrm{H2}}> 10^5$ \pcmc \ and $T_\mathrm{gas}\leq 110$ K (see Table~\ref{tab:groups}). From Figures~\ref{fig:SB_temp} and ~\ref{fig:SB_dens}, we see that for CS$_{\mathrm{comp2}}$, all the SB models, independently of the final temperature, gas density or CRIR, reproduce well the observed abundance. For SO$_{\mathrm{comp1b}}$, for the highest $\zeta/\zeta_0$ (corresponding to the outer GMCs where this SO component is detected), the best agreement occurs for $T_\mathrm{gas}=[100-150; 250-300]$ K and $n_\mathrm{H}\geq 10^6$ \pcmc, hence compatible with the physical conditions of the gas where both CS$_{\mathrm{comp2}}$ and SO$_{\mathrm{comp1b}}$ could be emitted.\\
\indent Therefore,  CS$_{\mathrm{comp2}}$ and SO$_{\mathrm{comp1b}}$ may probe the same dense warm quiescent gas within GMCs, in agreement with previous studies \citep[e.g.][]{leroy_alma_2015, holdship_energizing_2022, tanaka_2023}. CCS, on the other hand, probes shocked or post-shocked gas.

\subsubsection{Origin of compact species ($\sim3$ pc scales)}\label{subsec:compact-emission}

\begin{figure}
    \centering
    \includegraphics[width=1\linewidth]{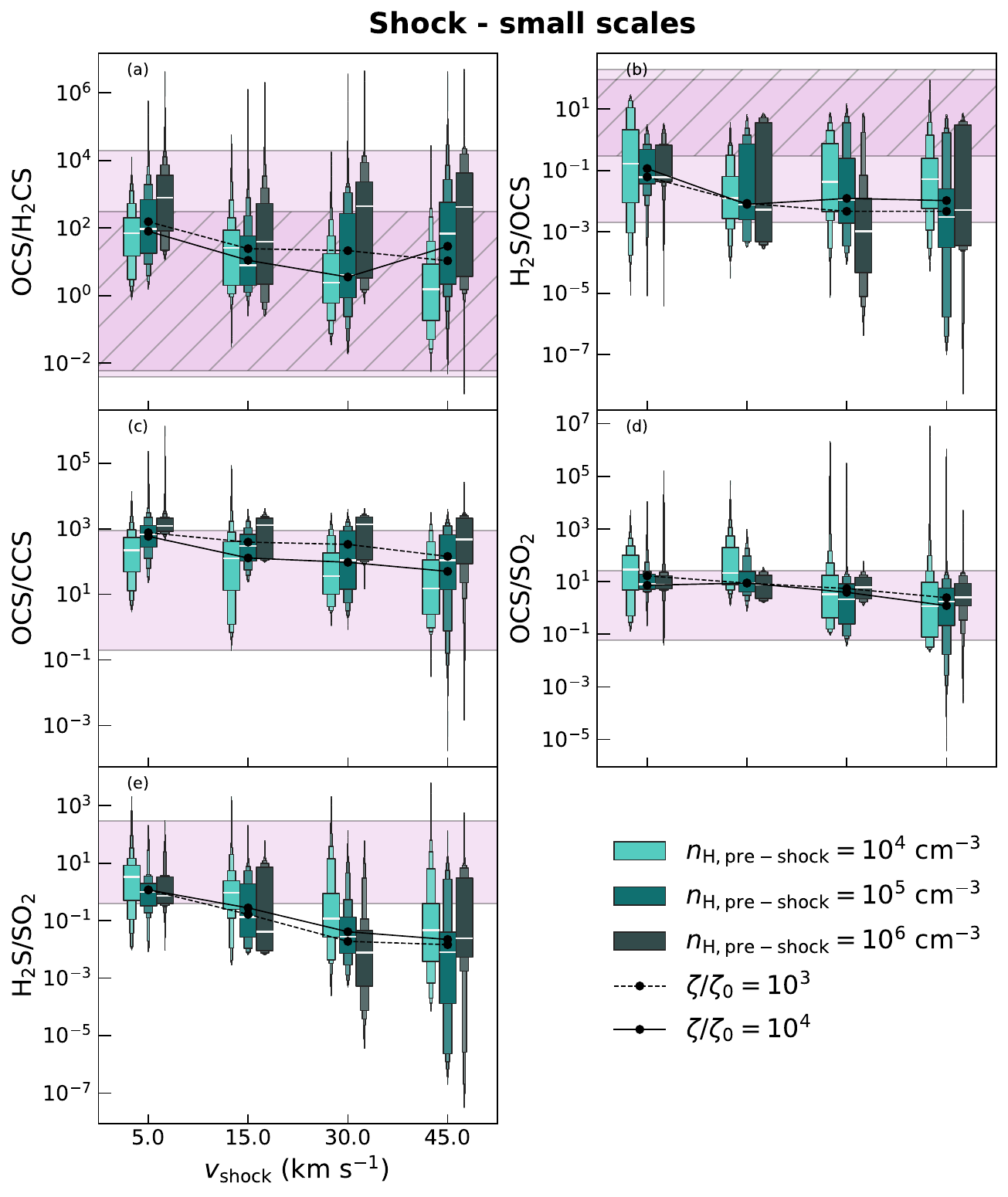}
    \caption{Same as Figure~\ref{fig:groupA} for the abundance ratios of species emitting on small (0.2$''$) scales and for shock models only. The y-axis is in log scale. For panels (a) and (b) we distinguish the observed ratios of the first components (OCS$_{\mathrm{comp1}}$/\ce{H2CS}$_{\mathrm{comp1}}$ and \ce{H2S}/OCS$_{\mathrm{comp1}}$; purple area) from the second components (OCS$_{\mathrm{comp2}}$/\ce{H2CS}$_{\mathrm{comp2}}$ and \ce{H2S}/OCS$_{\mathrm{comp2}}$; dashed purple area).}
    \label{fig:Ratio_groupC_shock}
\end{figure}

\begin{figure}
    \centering
    \includegraphics[width=1\linewidth]{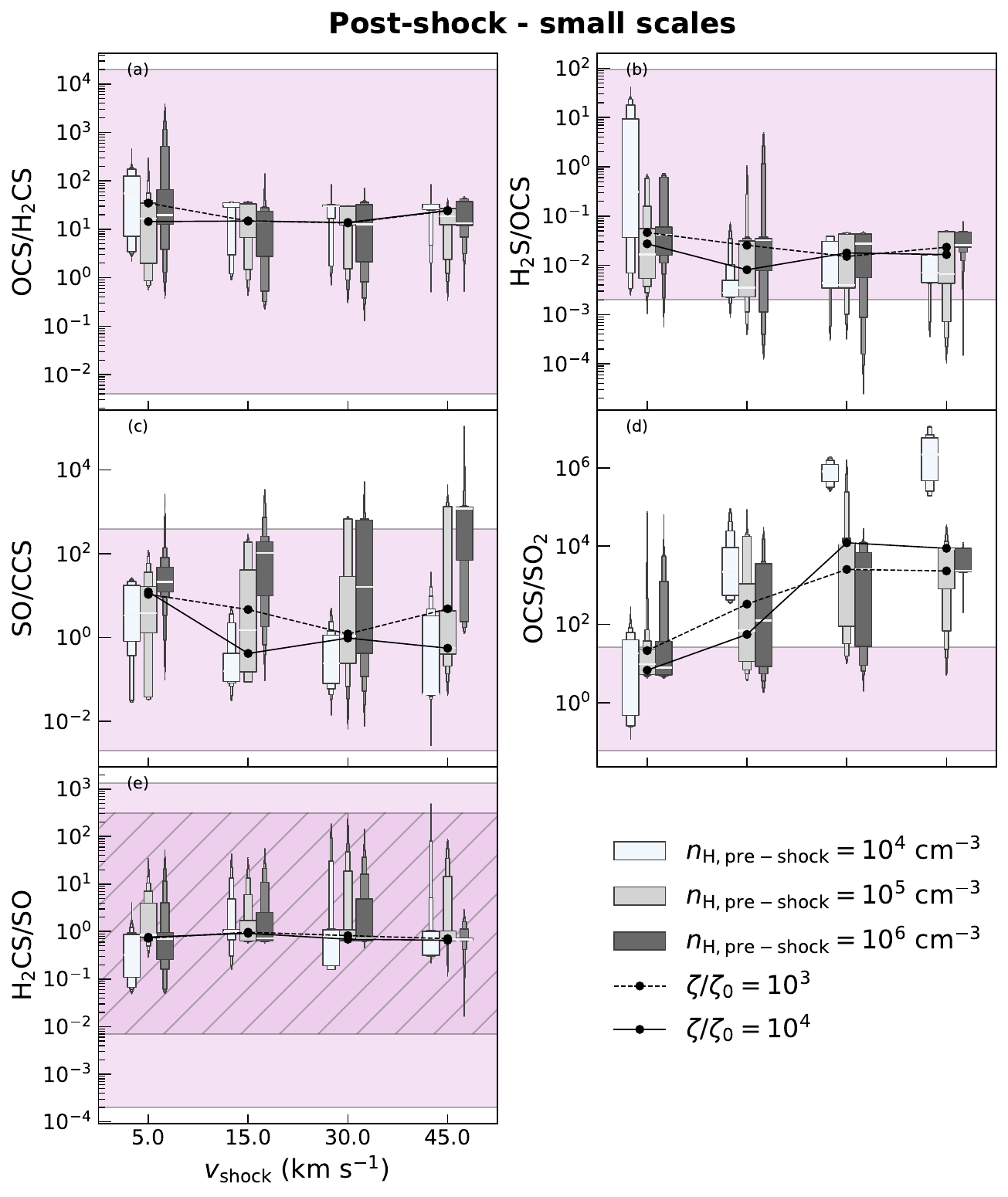}
    \caption{Same as Figure~\ref{fig:Ratio_groupC_shock} but for post-shock models. In panel (f) we differentiate the observed ratios \ce{H2CS}$_\mathrm{comp1}$/\ce{SO}$_\mathrm{comp1a}$ (light purple shaded area) and \ce{H2CS}$_\mathrm{comp2}$/\ce{SO}$_\mathrm{comp2}$ (dashed area).}
    \label{fig:Ratio_groupC_postshock}
\end{figure}

\begin{figure*}
    \centering
    \includegraphics[width=1\linewidth]{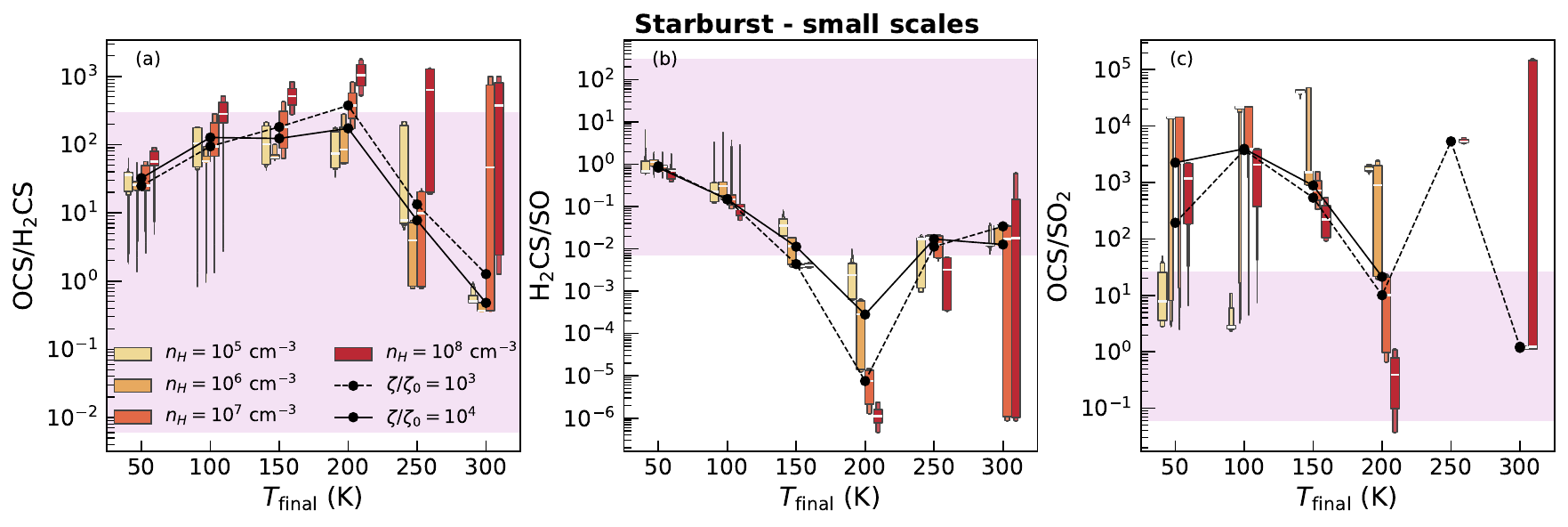}
    \caption{Same as Figure~\ref{fig:Ratio_groupC_shock} but for SB models. The abundances are as a function of the final gas temperature (x-axis) and density (colours).}
    \label{fig:Ratio_groupC_SB}
\end{figure*}

For species emitting at small scales (0.2$''$), we selected 8 ratios to test (see Table~\ref{tab:ratios}). We compare the various observed ratio ranges with the shock, post-shock and SB models in Figures~ \ref{fig:Ratio_groupC_shock}, \ref{fig:Ratio_groupC_postshock}, and ~\ref{fig:Ratio_groupC_SB}, respectively.

\vspace{-0.5cm}
\paragraph{\textbf{OCS$_{\mathrm{comp1}}$ and \ce{H2CS}$_{\mathrm{comp1}}$}:} 
Figures~\ref{fig:Ratio_groupC_shock}a and \ref{fig:Ratio_groupC_postshock}a show OCS/\ce{H2CS} as a function of the shock velocities. Both models agree with the observations, mainly due to the fact that the observed ratio spans several orders of magnitude. From Figures~\ref{fig:shock_vel} to ~\ref{fig:postshock_dens}, we see that the ratios do not depend on the shock velocity. For the pre-shock density, better agreement occurs for the shock models if $n_\mathrm{pre-shock}=10^4$ \pcmc \ for $\zeta/\zeta_0=10^3$ or up to $n_\mathrm{pre-shock}=10^5$ \pcmc \ for $\zeta/\zeta_0=10^4$ for OCS. For the post-shock models, only models with $n_\mathrm{pre-shock}\geq 10^5$ \pcmc \ reproduce the observations for both OCS and \ce{H2CS} for the highest CRIR. Overall, the models seem to support the observational conclusions (OCS probe shocks while \ce{H2CS} probes post-shocks; \citetalias{bouvier2024}), although we cannot distinguish between shock or post-shock origins.  

\vspace{-0.5cm}
\paragraph{\textbf{\ce{H2S} and  SO$_{\mathrm{comp2}}$:}} 
\ce{H2S} and SO$_{\mathrm{comp2}}$ were labelled as possible shock/post-shock tracers in \citetalias{bouvier2024}.
Figure~\ref{fig:Ratio_groupC_shock}b shows that the modelled ratios \ce{H2S}/OCS are in excellent agreement with the observations, independently of the parameters ($v_\mathrm{shock}, n_\mathrm{pre-shock}, \zeta/\zeta_0$). Looking further with Figures~\ref{fig:shock_vel} and ~\ref{fig:shock_dens}, we cannot constrain the shock velocity, but for the density, both abundances are simultaneously better reproduced for $n_\mathrm{pre-shock}=10^4$ \pcmc \ with $\zeta/\zeta_0=10^3$ and and for $n_\mathrm{pre-shock}\geq 10^5$ \pcmc \ with $\zeta/\zeta_0=10^4$ (condition relevant for the outer and inner GMCs, respectively). Because the physical conditions derived for SO$_{\mathrm{comp2}}$ are close to the one derived for \ce{H2CS}$_{\mathrm{comp2}}$ (Table~\ref{tab:groups}), we look at post-shock models for \ce{H2CS}$_{\mathrm{comp2}}$/SO$_{\mathrm{comp2}}$ (Figure ~\ref{fig:Ratio_groupC_postshock}e), where we see an excellent agreement with the observations. Figures ~\ref{fig:Ratio_groupC_postshock}e and ~\ref{fig:Ratio_groupC_SB}b, show the ratio \ce{H2CS}/SO in post-shock and SB models, respectively. Once again, all the post-shock models agree with the observations. For the SB models, the agreement with the observed range of the ratio occurs at all densities up to $T_\mathrm{final}=150$ K, which is consistent with the range of physical parameters derived ($T_\mathrm{gas}\geq 50$ K and $n_\mathrm{H2}\sim 10^5-10^7$ \pcmc; see Table~\ref{tab:groups}). Hence, we cannot conclude whether that SO$_{\mathrm{comp2}}$ comes from a post-shocked or a warm quiescent gas. Higher angular resolution observations may be necessary to disentangle the two possibilities.

\vspace{-0.5cm}
\paragraph{\textbf{CCS:}} 
To examine the shock scenario, we look at the ratio OCS$_{\mathrm{comp1}}$/CCS (Figure~\ref{fig:Ratio_groupC_shock}c): There is excellent agreement for $v_\mathrm{shock}\leq 30$ \kms \ and $n_\mathrm{pre-shock}\leq 10^5$ \pcmc \ and for $v_\mathrm{shock}\leq 45$ \kms \ at all pre-shock densities. If both species probe the same type of shock, it is rather for $n_\mathrm{pre-shock}= 10^4$ \pcmc \ with $\zeta/\zeta_0=10^3$ and $n_\mathrm{pre-shock}\leq 10^5$ \pcmc \ with $\zeta/\zeta_0=10^4$ (see Figure~\ref{fig:shock_dens}). We test the post-shock scenario using the SO$_{\mathrm{comp2}}$/CCS ratio, assuming SO$_{\mathrm{comp2}}$ probes post-shocked gas. SO$_{\mathrm{comp2}}$ being present only towards the inner GMCs, we consider only models with $\zeta/\zeta_0=10^4$. Figure~\ref{fig:Ratio_groupC_postshock}c show a good agreement with the observations at all pre-shock densities if $v_\mathrm{shock}\leq 30$ \kms.  Figures~\ref{fig:postshock_vel} and ~\ref{fig:postshock_dens} show that both species probe the same post-shocked gas only if $v_\mathrm{shock}=5$ \kms \ and $n_\mathrm{pre-shock}=10^5$ \pcmc, similar conditions found for  \ce{H2CS}$_{\mathrm{comp1}}$. However, from the temperature and density ranges derived from the observations (see Table~\ref{tab:groups}), SO$_{\mathrm{comp2}}$ and \ce{H2CS}$_{\mathrm{comp1}}$ do not probe the same physical conditions. Hence, SO$_{\mathrm{comp2}}$, CCS, and  \ce{H2CS}$_{\mathrm{comp1}}$ could probe post-shocked gas, but not necessarily from the same type of shock.

\vspace{-0.5cm}
\paragraph{\textbf{SO$_{\mathrm{comp1a}}$:}} 
Figures~\ref{fig:low-dens-SB} and ~\ref{fig:low-dens-SB-temp} show that the abundances for $\zeta/\zeta_0=10^3$ (conditions for the outer GMCs, where this SO component is detected) are in agreement with the observations primarily for $T_\mathrm{final}=150-250$ K and $n_\mathrm{H}=10^4$ \pcmc. Our models thus support SO$_{\mathrm{comp1a}}$ as probing a warm dense quiescent gas. 
\indent We use the ratio OCS$_{\mathrm{comp1}}$/SO$_{\mathrm{comp1a}}$ to check the initial hypothesis that the two species are not probing the same gas. Figure~\ref{fig:lowdensratio}b shows that the ratio in SB models reproduces well the observations. However, Figure~\ref{fig:low-dens-SB-temp} shows that the best agreement with the observed abundances of OCS and SO$_{\mathrm{comp1a}}$ does not really occur around the same temperatures. Hence, the two species do not trace the same quiescent warm gas. In addition, Figure ~\ref{fig:Ratio_groupC_postshock}e shows the \ce{H2CS}$_{\mathrm{comp1}}$/SO$_{\mathrm{comp1a}}$ ratio in post-shock models: the models are in good agreement with the observations, independent of the shock velocity, the pre-shock density or the CRIR (see also Figures~\ref{fig:postshock_vel} and ~\ref{fig:postshock_dens}). Hence, we cannot exclude that SO$_{\mathrm{comp1a}}$ could be also compatible with a post-shocked gas origin.

\vspace{-0.5cm}
\paragraph{\textbf{OCS$_{\mathrm{comp2}}$, \ce{H2CS}$_{\mathrm{comp2}}$, and  SO$_2$:}} 
We consider only the models with $\zeta/\zeta_0=10^4$, since these components are only detected in the inner GMCs.  
Results for the OCS$_{\mathrm{comp2}}$/\ce{SO2} ratio are presented in Figures~\ref{fig:Ratio_groupC_shock}d, \ref{fig:Ratio_groupC_postshock}d and \ref{fig:Ratio_groupC_SB}c. SB models reproduce the observations with $T_{\mathrm{final}}\leq 100$ K and $n_{\mathrm{H}}=10^5$ \pcmc \ or $T_{\mathrm{final}}= 200$ K and $n_{\mathrm{H}}\geq10^7$ \pcmc, in agreement with what can be found when comparing their individual abundances ( Figures~\ref{fig:SB_temp} and ~\ref{fig:SB_dens}). Shock models also reproduce well the observations, but Figures~\ref{fig:shock_vel} and ~\ref{fig:shock_dens} show that this does not occur under the same conditions. The same argument is valid in the case of post-shock models (Figures~\ref{fig:postshock_vel} and ~\ref{fig:postshock_dens}). Hence, a common shock or post-shock origin is unlikely. We can test a shock origin comparing OCS$_{\mathrm{comp2}}$ and \ce{SO2} with \ce{H2S}, which we labelled as a shock tracer. The ratios \ce{H2S}/OCS$_{\mathrm{comp2}}$ and \ce{H2S}/\ce{SO2} are presented in Figure~\ref{fig:Ratio_groupC_shock}b and Figure~\ref{fig:Ratio_groupC_shock}e, respectively. There is a clear disagreement with the observations for \ce{H2S}/OCS$_{\mathrm{comp2}}$, as the ratio is mostly underestimated. For \ce{H2S}/\ce{SO2}, shock models best agree with the observations only if $v_\mathrm{shock}=5$ \kms. However, abundances of \ce{SO2} and \ce{H2S} simultaneously reproduce the observations for $v_\mathrm{shock}\leq 15$ \kms \ and $n_\mathrm{pre-shock}=10^4$ \pcmc \ as shown in Figures~\ref{fig:shock_vel} and \ref{fig:shock_dens}. At these compact scales, where pSSCs are present, a density $\geq 10^6$ \pcmc \ is expected (e.g. \citealt{rico-villas_super_2020}). Hence, OCS$_{\mathrm{comp2}}$ and \ce{SO2} probably do not probe shocks. A hot quiescent gas origin due to the presence of pSSCs is the most likely scenario. \\
\indent The ratio OCS$_{\mathrm{comp2}}$/\ce{H2CS}$_{\mathrm{comp2}}$ (Figure~\ref{fig:Ratio_groupC_SB}a), agrees with the observations at all temperatures for $n_\mathrm{H}\leq 10^7$ \pcmc. A common origin for these three components is possible (see Figures~\ref{fig:SB_temp} and ~\ref{fig:SB_dens}). 
On the other hand, both a shock and post-shock origin cannot be excluded for \ce{H2CS}$_{\mathrm{comp2}}$: the ratio \ce{H2CS}$_{\mathrm{comp2}}$/SO$_{\mathrm{comp2}}$ matches well the observations at all shock velocities and pre-shock densities (Figure~\ref{fig:Ratio_groupC_postshock}e), and their individual abundances can be reproduced under the same conditions ($v_\mathrm{shock}=5$ \kms \ and $n_\mathrm{pre-shock}=10^5-10^6$ \pcmc; see Figures~\ref{fig:postshock_vel} and \ref{fig:postshock_dens}). Finally, Section~\ref{subsec:tools} showed that shock models statistically better reproduce the observations of \ce{H2CS}$_{\mathrm{comp2}}$. Hence, we cannot conclude whether \ce{H2CS}$_{\mathrm{comp2}}$ probes shocked, post-shocked or warm quiescent gas, although an origin from the pSSCs would be in agreement with previous studies \citep{krieger_molecular_2020a}. Higher angular resolution observations are needed to solve the question.

\subsection{The sulphur depletion level in the CMZ of NGC\,253?}

\begin{figure*}
    \centering
    \includegraphics[width=1\linewidth]{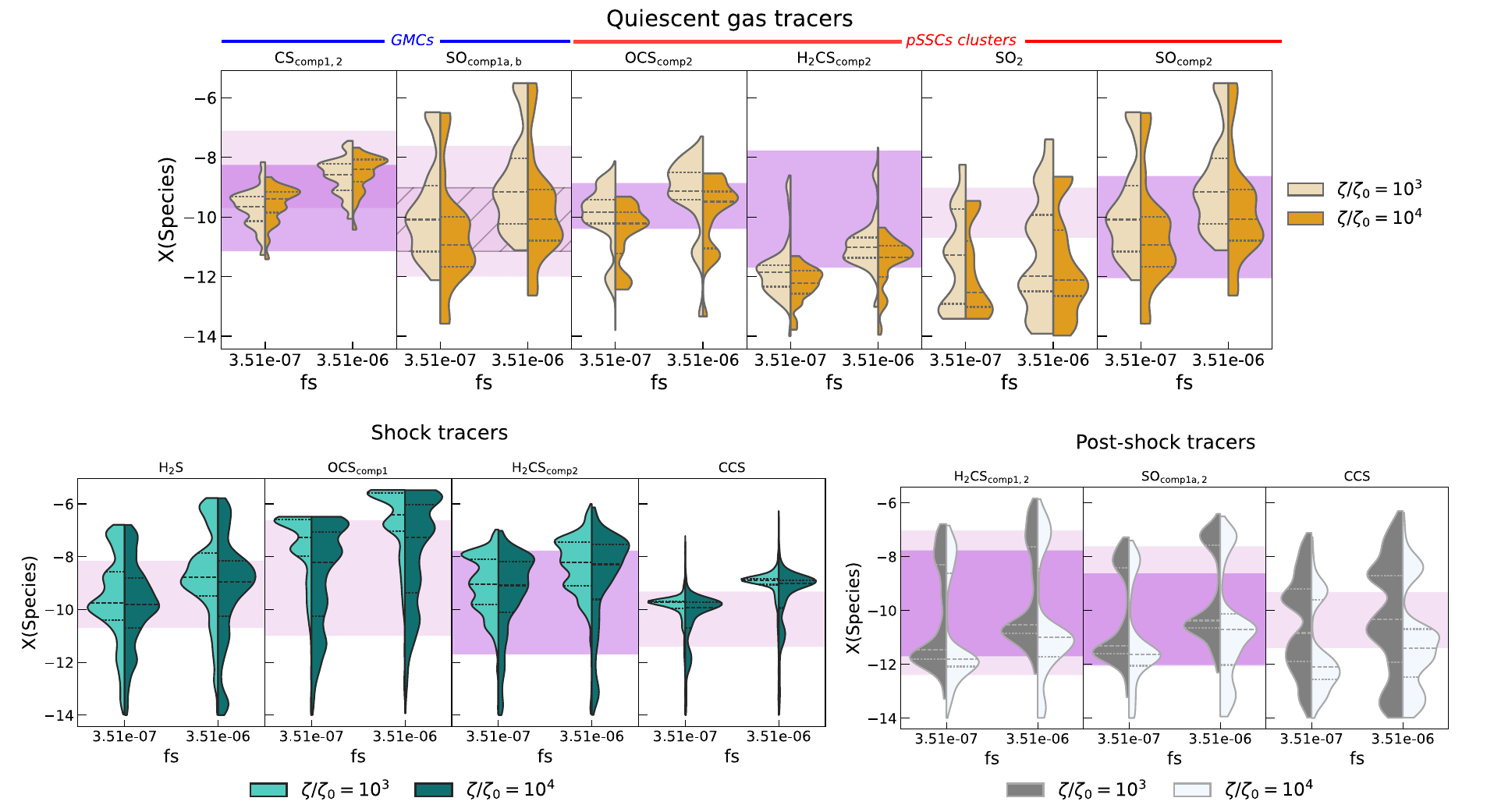}
    \caption{Violin plots of the fractional abundances in SB (oranges, Top), shock (cyans, Middle) and post-shock (greys, Bottom) models as a function of the sulphur abundance (x-axis) and $\zeta/\zeta_0$ (shades of colours). The interquartiles of the data (representing the middle 50\% of the distribution) are shown. The observed abundances are shown in a horizontal shaded purple area. For species presenting a second component, the shaded area is darker. We distinguish SO$_{\mathrm{comp1a}}$ (light purple) from SO$_{\mathrm{comp1b}}$ (dashed area). }
    \label{fig:fs}
\end{figure*}

In dense Galactic star-forming regions, the sulphur abundance accounts between 1 and 10\% of the  Solar value \citep[e.g.][]{Ruffle1999, vastel_sulphur_2018, holdship_sulfur_2019,  bouscasse_sulphur-rich_2022,fuente_gas_2023, Tang2024}. On the other hand, more diffuse and highly ionised regions show much less or essentially no depletion \citep[e.g.][]{goicoechea_low_2006, Neufeld2015, riviere-marichalar_abundances_2019, goicoechea_bottlenecks_2021, Fuente2024}. 
As the CMZ of NGC 253 hosts intense high-mass star formation activity, we investigate whether we can constrain the level of sulphur depletion within these extreme starburst conditions. 

To investigate the level of sulphur depletion in the CMZ of NGC\,253, we consider the models with different sulphur fractional abundances, coupled with small (fs=$3.51\times10^{-6}$) or large (fs=$3.51\times10^{-7}$) depletion (see Figure~\ref{fig:fs}). For each model type, we only consider the species and components for which we have concluded that the model could explain their emission. 
From Figure~\ref{fig:fs}, we see that:\\
$-$ In GMC-like gas tracers, while CS$_\mathrm{comp1}$ is more consistent with a low depletion case, CS$_\mathrm{comp2}$ is more consistent with the heavier depletion case. CS$_\mathrm{comp2}$ likely probes denser gas compared to CS$_\mathrm{comp1}$ (see Table~\ref{tab:groups}). This result is consistent with previous studies that found that sulphur depletion increases with increasing density \citep[e.g.][]{hily-blant_sulfur_2022}.\\
$-$ For the tracers of more dense, compact and hot gas the lower depletion is favoured in the case of \ce{H2CS}$_\mathrm{comp2}$. Neither option seems to agree with the measured \ce{SO2} abundances, although \ce{SO2} seems to be more produced in the lowest depletion case. This would imply that, at the scale of pSSCs, a lower sulphur depletion is favoured, consistent with studies of high-mass star-forming regions \citep[e.g.][]{fuente_gas_2023, Fuente2024}. However, \ce{H2CS}$_\mathrm{comp2}$ may not necessarily probe the quiescent hot gas from pSSC groups (see Section~\ref{subsec:compact-emission}).\\
$-$ For the shock tracers, only \ce{OCS}$_\mathrm{comp1}$, CCS, and to some extent \ce{H2CS}$_\mathrm{comp2}$ favour  higher depletion. If CCS preferentially probes post-shocked gas, then the lower depletion case is more viable. This result seems surprising since shocks in Galactic regions appear to have a lower sulphur depletion compared to cold molecular clouds, as sulphur is expected to be efficiently released from the grains through sputtering \citep[e.g.][]{bachiller_shock_1997, anderson_new_2013, holdship_sulfur_2019}. Since we could not put any constraints on the type of shocks probed by OCS or CCS, we are not able to constrain the sulphur depletion level for these species.

\indent Our results indicate that different regions within the GMCs have different sulphur depletion levels: it is higher in the densest cold/warm regions of the GMCs and in the shocked regions than in the less dense part of the GMCs and in the hot gas close to the pSSCs.  
However, our results mostly rely on species for which the origin of emission is the least certain. We also did not study all the sulphur-bearing species detected towards the CMZ of NGC\,253 (such as the isotopologues of CS, NS, \ce{HCS+}, \ce{CH3SH}, \ce{SO+}; \citealt{martin_alchemi_2021}), since we focussed on those studied observationally from \citetalias{bouvier2024}. The inclusion of all the detected sulphur-bearing species as well as a higher angular resolution are needed in order to confirm our results.

\section{Conclusions}\label{sec:conclusions}
In this work, we investigate the origin of the emission from seven sulphur-bearing molecular species detected in the CMZ of the starburst galaxy NGC\,253. 
We employed a chemical model to create static warm clouds and C-type shocks under the physical conditions representative of the CMZ of NGC\,253, and compared with observations performed in \citetalias{bouvier2024}. 
Our main findings are summarised as follows.

\begin{enumerate}
    \item We could confirm observationally-based hypotheses from \citetalias{bouvier2024}, and constrain some of the most compact emission of sulphur-bearing species. We found that these species cannot be reproduced by a single physical model.\\
        $-$ At GMC-like scales ($\sim 30$ pc), CS probes quiescent dense gas. Part of the CS emission also probes a more compact region within the GMCs ($\sim 10$ pc), similar to SO emission. \\
        $-$ At the smallest scales ($\sim 3$ pc), while \ce{H2S}, OCS are consistent with shocked gas, post-shocked gas is favoured for \ce{H2CS}. Compact SO emission probes dense quiescent gas, although a post-shock origin could not be excluded. \\
        $-$ The highest-excitation emission of OCS, \ce{SO2} and \ce{H2CS} are probing the hot gas linked to pSSCs, although \ce{H2CS} may be related to shocked gas. Higher angular resolution is needed to conclude. \\
        $-$ The emitting region from which CCS originates is not clear, but a shocked or a post-shocked gas origin is favoured.

        \item Under the extreme conditions of the CMZ of NGC\,253, and provided that the abundances are well constrained, we found that most sulphur-bearing species investigated in this work may be used to distinguish between quiescent, shocked and post-shocked gas. We found that distinguishing post-shocked from other processes is not straightforward, as warm chemistry prevails over freeze-out processes during this phase.

        \item Overall, the initial temperature, CRIR and shock velocity had the least impact when it comes to comparing models and observations. In the case of the CRIR, this is likely due to the fact that we tested two high values. CS is the least impacted by the various physical parameters and can be used as a reference standard when paired with other species sensitive to CRIR, in agreement with \citet{Dutkowska2025}. 

        \item We investigated the sulphur depletion level and found that different levels are favoured for different species. This could imply that we are sensitive to different regions within the GMCs  presenting different depletion levels, even using unresolved observations. We found that close to the pSSCs, the best-matching depletion level is the lowest, similar to Galactic highly ionised regions. Higher angular resolution observations are needed to confirm this result. 
\end{enumerate}

In extragalactic observations,
observing different species allow us to probe different unresolved ISM components \citep[e.g.][]{Viti2017}.
This work showed that combining observations and chemical modelling of sulphur-bearing species is an efficient technique for studying extragalactic star-forming regions. More specifically, we found that these species can be effective in distinguishing between quiescent and shocked gas associated with ongoing star-formation, and that they are thus powerful tools that can be used to study other extragalactic starburst environments.

\begin{acknowledgements}
M.B., S.V., and K.M.D. acknowledge the support from the European Research Council (ERC) Advanced Grant MOPPEX 833460.     

\end{acknowledgements}

%
  \bibliographystyle{bibtex/aa} 
  \bibliography{main} 

\onecolumn
\FloatBarrier
\begin{appendix}

\section{Abundance calculations}\label{subsec:abun}

To calculate the observed abundances (with respect to \ce{H2}), we used the column density range for each sulphur-bearing species obtained through non-LTE calculations (except for CCS, \ce{SO2} and the high-J lines of CCS for which we used the LTE results) in \citetalias{bouvier2024}. For the \ce{H2} column densities, we used the range derived by \cite{Behrens2024} through the multi-transition analysis of HCN and HNC. Although they used a 50 pc-size region in their analysis, the derived N(\ce{H2}) values were in in agreement with those derived in previous ALCHEMI studies (see their Table 3). They derived two ranges of N(\ce{H2}) values across the CMZ of NGC\,253, one for regions located within 100 pc around the kinematic centre of the CMZ (inner region;  N(\ce{H2}) ranging between $6.3\times10^{23}$ \pcms \ and $5.0\times10^{25}$ \pcms) and one for regions located beyond 100 pc (outer region; N(\ce{H2}) ranging between $3.2\times10^{22}$ \pcms and $3.2\times10^{23}$ \pcms). The inner region includes the innermost GMCs (3 to 7) while the outer regions include the outermost GMCs (10, 9a, 8a, 1a, 1b, and 2b). 
Since the derived \ce{H2} column densities are those derived at scales larger than the ALCHEMI beam size, we apply a beam-filling factor $ff$ for each region of emission: $ff=\theta_s^2/(\theta_s^2+\theta_b^2)$ where $\theta_b$ is the beam size of $1.6''$ and $\theta_s$ is the source size of $1.6''$, $0.6''$, and $0.2''$. This results in beam filling factors of 0.5, 0.12, and 0.015, respectively. For CCS, \ce{SO2} and OCS$_{\mathrm{comp2}}$, the derived column densities in \citetalias{bouvier2024} are beam-averaged column densities, hence we also apply the beam filling factor to the column densities. 

\section{Low density models}\label{appdx:low_dens_models}

\begin{figure}[H]
    \centering
    \includegraphics[width=0.8\linewidth]{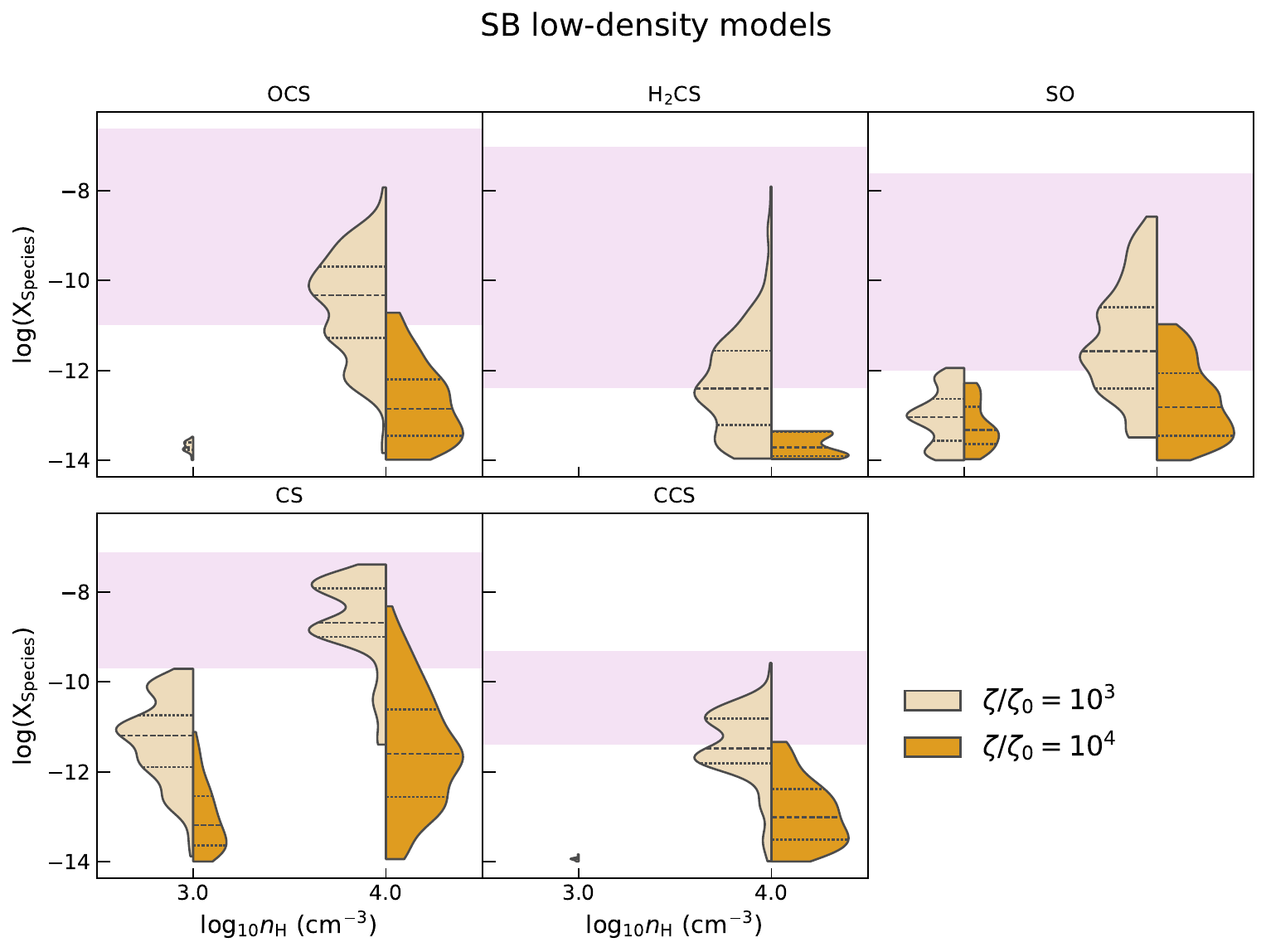}
    \caption{Violin plots of the fractional abundances in SB models as a function of the final gas density (x-axis) and $\zeta/\zeta_0$ (shaded colors). The interquartiles of the data are shown. The observed abundances are shown as a horizontal shaded purple area. The components considered for the observational abundances are OCS$_\mathrm{comp1}$, \ce{H2CS}$_\mathrm{comp1}$, SO$_\mathrm{comp1a}$, CS$_\mathrm{comp1}$, and CCS.}
    \label{fig:low-dens-SB}
\end{figure}

\begin{figure}
\centering
\includegraphics[width=0.8\linewidth]{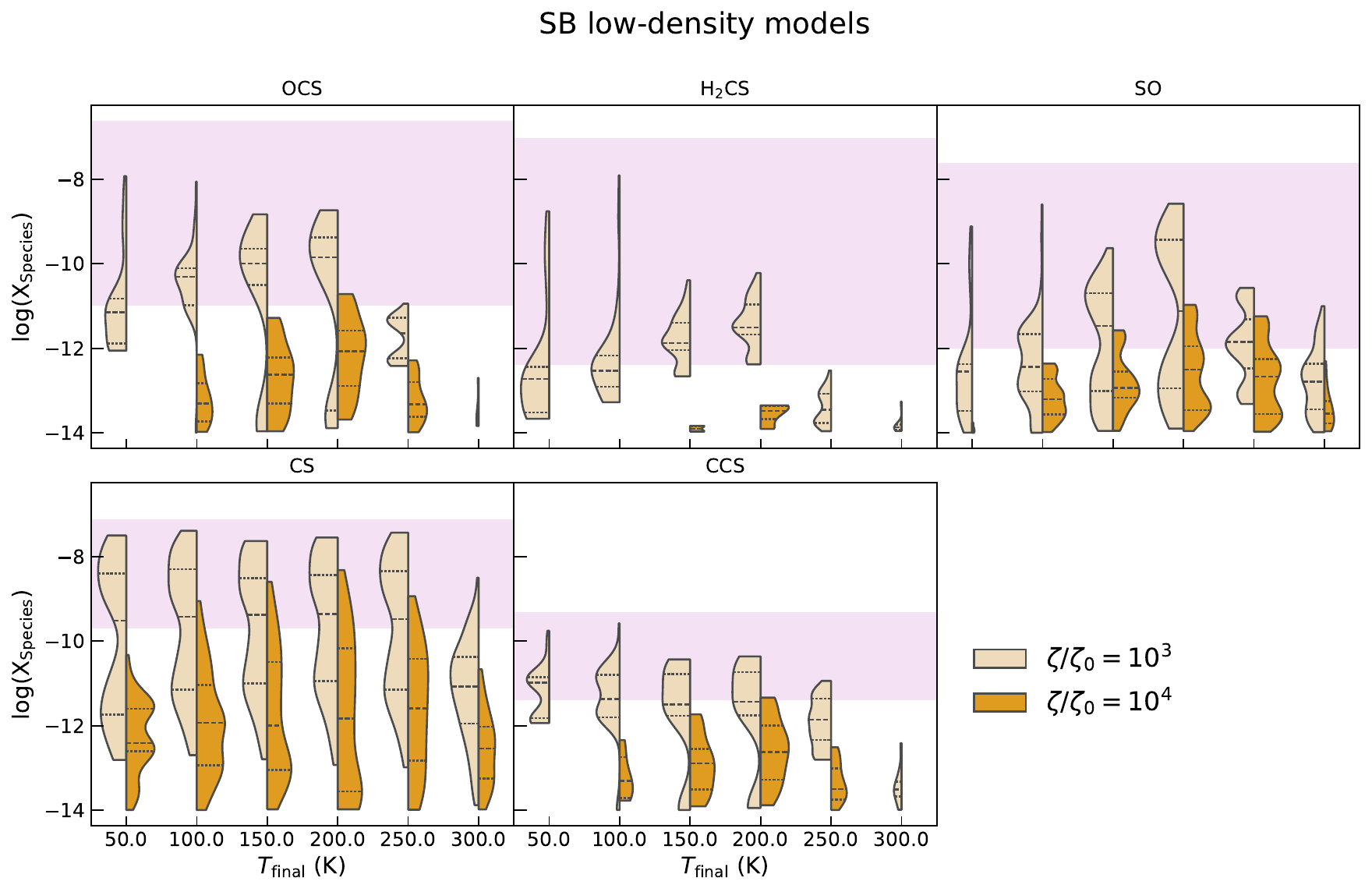}
\caption{Same as Figure~\ref{fig:low-dens-SB} as a function of the final temperature (x-axis).}
\label{fig:low-dens-SB-temp}
\end{figure}

\begin{figure}
    \centering
    \includegraphics[width=0.7\linewidth]{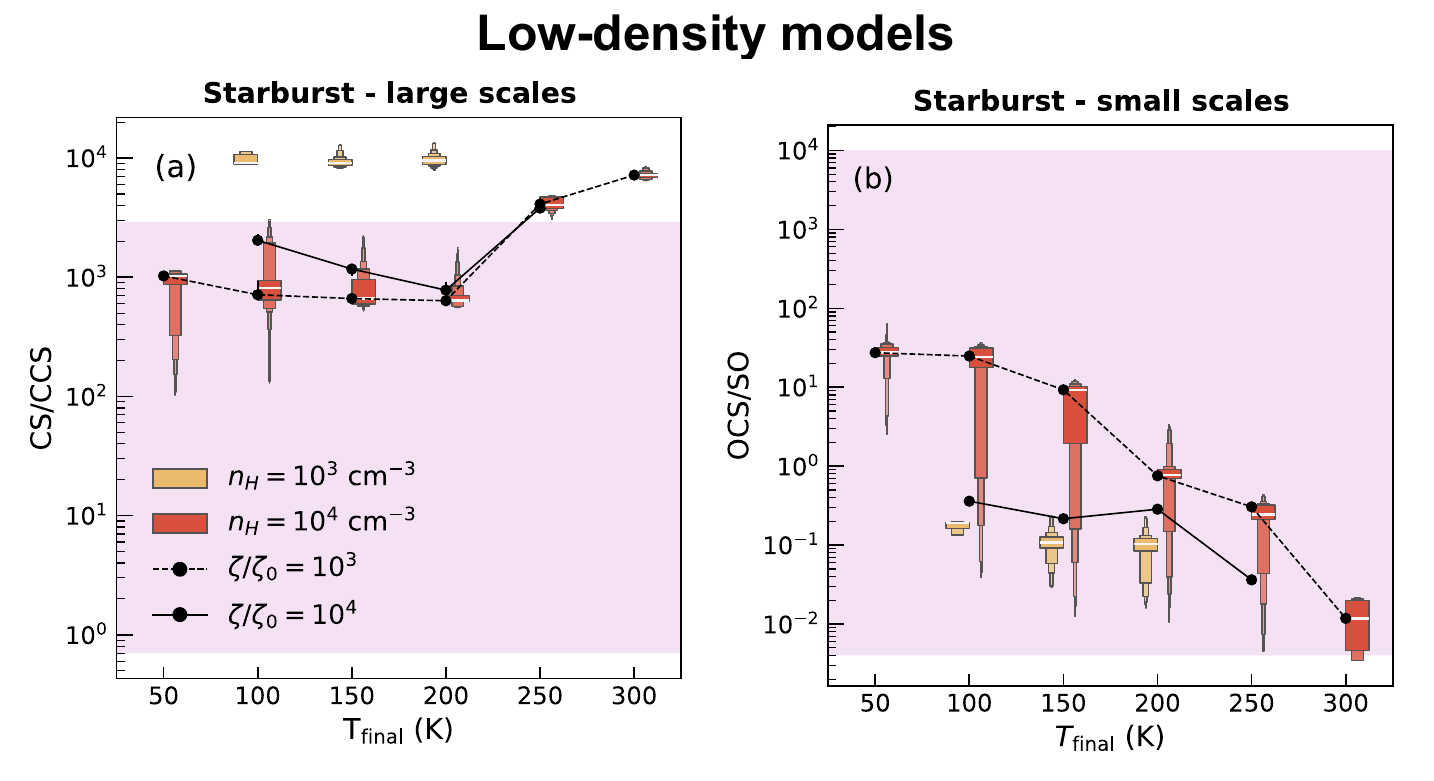}
    \caption{Boxen plots of the abundance ratio of species emitting on large (1.6$''$; left) and small ($0.2''$; right) scales for the SB models as a function of the final gas temperature (x-axis) and final density (shades of colours). The width of the boxes correspond for each percentile with the median (Q2, 50th percentile) highlighted by the white horizontal line. The extent of the boxen plots covers the full range of data. For all models, the modelled median ratio as a function of $\zeta/\zeta_0$ is indicated by the dashed ($\zeta/\zeta_0=10^3$) and solid ($\zeta/\zeta_0=10^4$) lines. The ratio range derived from the observations are shown with the purple shaded area.}
    \label{fig:lowdensratio}
\end{figure}

We ran additional SB models with final densities of $n_\mathrm{H}=10^3-10^4$ \pcmc to match with the derived densities of CS$_\mathrm{comp1}$, SO$_\mathrm{comp1a}$, OCS$_\mathrm{comp1}$, \ce{H2CS}$_\mathrm{comp1}$ in some of the GMCs \citep{bouvier2024}. We compare the modelled abundances of these species at these low densities in Figures~\ref{fig:low-dens-SB} and \ref{fig:low-dens-SB-temp}. We also include the result for CCS, since this species could arise from a quiescent molecular cloud-like gas \citep[e.g.][]{suzuki_1992, rathborne_2008, hirota_2009, roy_2011, Tatematsu2010, tatematsu2014b, seo_2019, koley_2022, Worthen2025}.
Figure~\ref{fig:lowdensratio}a shows the ratio of species emitting at large scales with the low-density models, as these species emit on GMC-scales, where such densities are expected.

\FloatBarrier
\section{Effects from the physical conditions}\label{appdx:params_dependence}
In this section, we look at how different input parameters influence the fractional abundances of each sulphur-bearing species. In all of the models, we investigate the effect of high cosmic ray ionization rates and initial temperatures. In SB models, we investigate the effect of the final gas temperature and gas density, while in shock and post-shocked models we investigate the effects of the shock velocity and pre-shock density.

\paragraph{Initial temperature ($T_{\mathrm{init}}$):}

The choice of the initial temperature, i.e. the temperature at which the cloud collapses during Stage 1, does not affect the abundances of the sulphur-bearing species in the SB models. For the shock models, an increase from $T_{\mathrm{init}}=15$ K to $T_{\mathrm{init}}=20$ K leads to a decrease in the maximum abundances of \ce{H2S} and SO, up to one order of magnitude in the case of \ce{H2S}. The median modelled abundances of \ce{H2S}, \ce{H2CS}, SO, and CS also decrease with an increasing  $T_{\mathrm{init}}$. On the other hand, the maximum abundance of OCS increases from $T_{\mathrm{init}}=15$ K  to $T_{\mathrm{init}}=20-25$ K and the median and maximum abundances of \ce{SO2} slightly increase (less than 1 order of magnitude) for $T_{\mathrm{init}}=20$ K. Finally, the maximum abundances of \ce{H2CS}, CS, and CCS remain relatively constant with $T_{\mathrm{init}}$. In the post-shock models, the behaviour of the abundances for each species follows that of the shock models. Overall, the initial temperature does not provide further constraints when comparing the modelled and observed abundances.

\paragraph{High cosmic-ray ionization rates ($\zeta/\zeta_0$):}\label{par:crir}
In our models, we considered only CRIR of $10^3$ and $10^4$ times the galactic CRIR, which are values derived in the outer and inner part of the CMZ of NGC\,253 \citep{behrens_tracing_2022,Behrens2024}. We investigate here the effect of increasing $\zeta/\zeta_0$ on the modelled abundances. 
In the SB models, the median abundances decrease with the increasing CRIR for most species, except CS and CCS, for which the median abundances are slightly increasing or increasing by about one order of magnitude, respectively. In shock models, median abundances for OCS are clearly decreasing for all $T_{\mathrm{init}}$, while the effect is reduced at $T_{\mathrm{init}}=20-25K$ for \ce{H2S} and SO. The median abundances do not show large variations for the rest of the species. Similarly, as with $T_{\mathrm{init}}$, the CRIR does not provide further constraints when comparing the modelled and observed abundances.

\paragraph{Gas temperature ($T_{\mathrm{gas}}$):}

\begin{figure}
    \centering
    \includegraphics[width=1\linewidth]{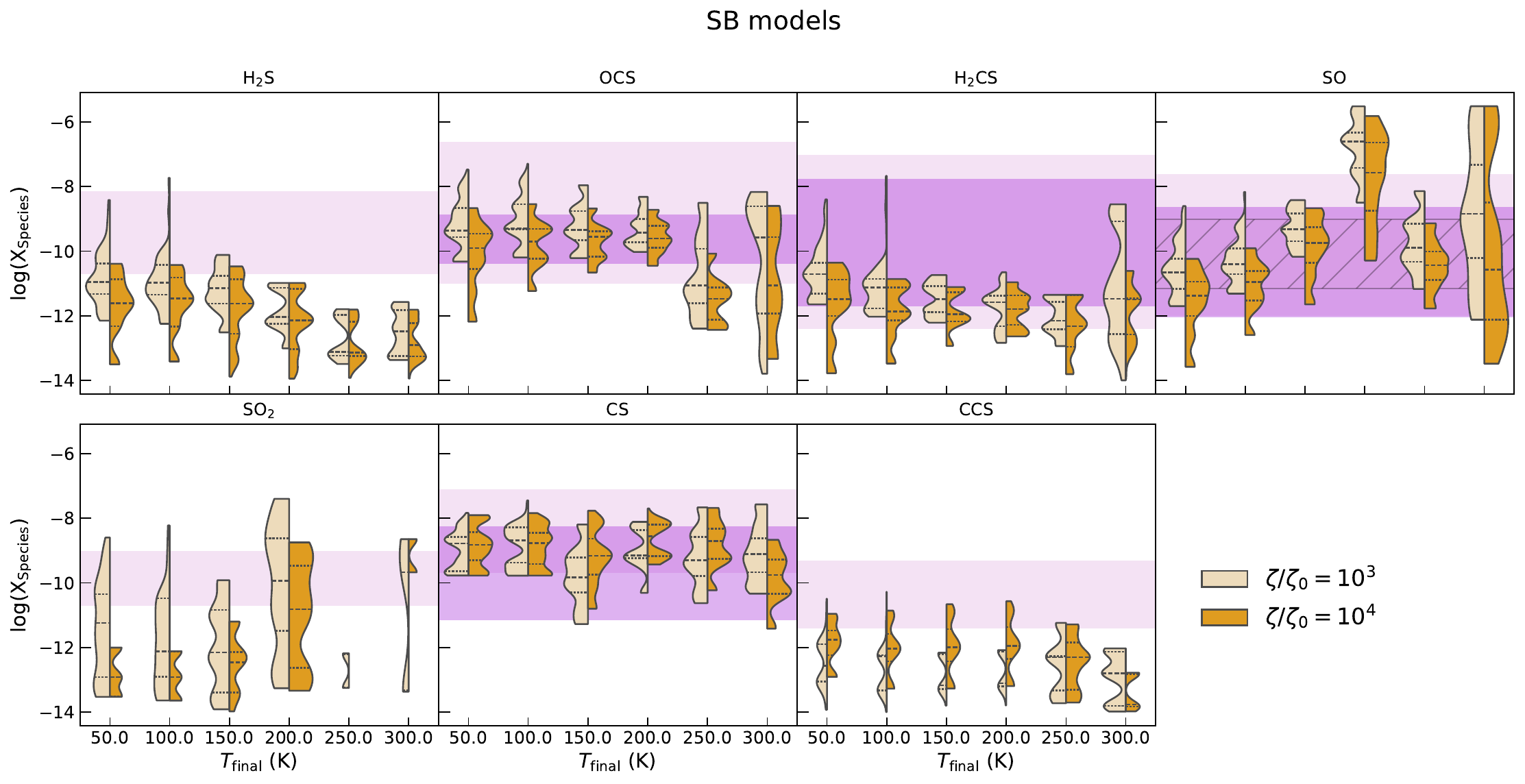}
    \caption{Violin plots of the fractional abundances in SB models as a function of the final gas temperature (x-axis) and $\zeta/\zeta_0$ (shaded colors). The interquartiles of the data are shown. The observed abundances are shown as horizontal shaded purple areas. For species presenting a second component, the shaded area is darker. In the SO case (see text), we distinguish SO$_{\mathrm{comp1a}}$ (light purple) from SO$_{\mathrm{comp1b}}$ (dashed area). }
    \label{fig:SB_temp}
\end{figure}

We investigate the effect of the final temperature reached in the SB models on the modelled abundances of sulphur-bearing species in Figure~\ref{fig:SB_temp}. 
The maximum abundances of \ce{H2S} decrease with increasing final temperature. For $T_{\mathrm{final}}> 200$ K, the modelled abundances do not agree with the observed abundances. 
The maximum abundances of OCS and \ce{H2CS} decrease with increasing $T_{\mathrm{final}}$, and increase again at $T_{\mathrm{final}}=300$ K, in particular for $\zeta/\zeta_0=10^3$ in the case of \ce{H2CS}. The maximum abundances of SO and \ce{SO2} show similar behaviour, with an increase up to $T_{\mathrm{final}}=200$K and a drop in abundance at $T_{\mathrm{final}}=250$K. Finally, the abundances of CS and CCS do not vary with respect to $T_{\mathrm{final}}$, except for $T_{\mathrm{final}}\geq 250$K in the case of CCS, with a drop in the maximum abundance. All the models stay generally coherent with the observations, except for CCS for $T_{\mathrm{final}}=300$K. The modelled abundances of \ce{SO2} agree with observations for $T_{\mathrm{final}}\leq150$K only for $\zeta/\zeta_0=10^3$ and for both CRIR values for $T_{\mathrm{final}}=200$ or 300 K. 

\paragraph{Final density ($n_{\mathrm{H}}$):}

\begin{figure*}
    \centering
    \includegraphics[width=1\linewidth]{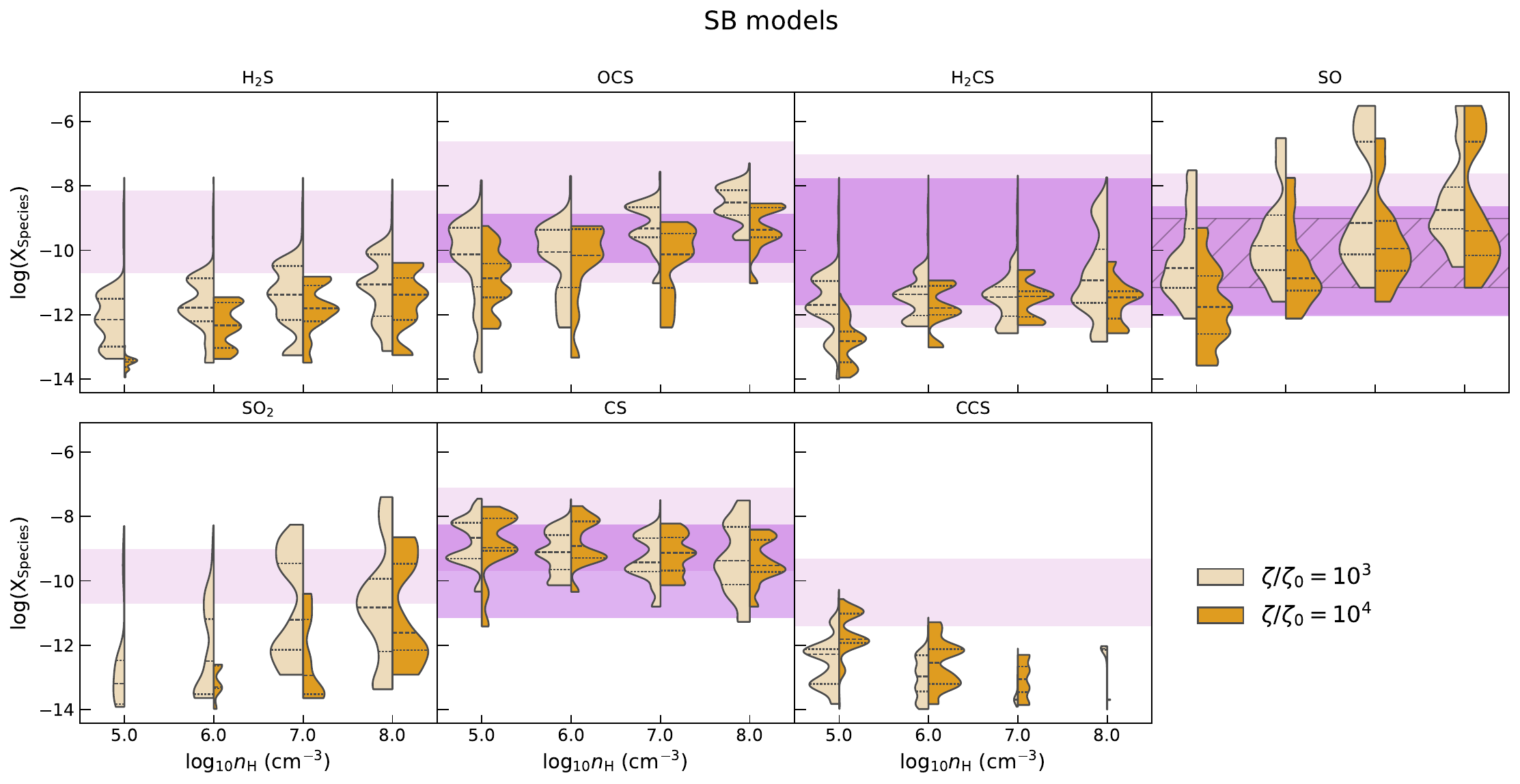}
    \caption{Same as Fig.~\ref{fig:SB_temp} but as a function of the final density (x-axis) and $\zeta/\zeta_0$ (shaded colours).}
    \label{fig:SB_dens}
\end{figure*}

Figure.~\ref{fig:SB_dens} shows the dependence of the abundances as a function of the final density, $n_\mathrm{H}$. Whilst the maximum abundances of SO and \ce{SO2} increase with increasing density, those for CCS decrease. The final densities do not affect the maximum abundances for the rest of the species. From the previous results, CS$_{\mathrm{comp1}}$ favours the SB models. Models with $n_\mathrm{H}\geq 10^6$\pcmc better agree with the observations.
For OCS, the best agreement with the observations for both CRIR values are for the intermediate values of $n_\mathrm{H}$ ($n_\mathrm{H}=10^6-10^7$ \pcmc).
In the case of \ce{SO2}, the best agreement with the observations are for the highest densities ($n_\mathrm{H}=10^7-10^8$ \pcmc).
Finally, \ce{H2S} and CCS do not favour the SB models, but the best agreement is for models with $n_\mathrm{H}\geq 10^7$\pcmc and $n_\mathrm{H}\leq 10^6$\pcmc, for \ce{H2S} and CCS, respectively. No additional constraints can be derived for the rest of the species.

\paragraph{Shock velocity ($v_{\mathrm{shock}}$):}

\begin{figure}
    \centering
    \includegraphics[width=1\linewidth]{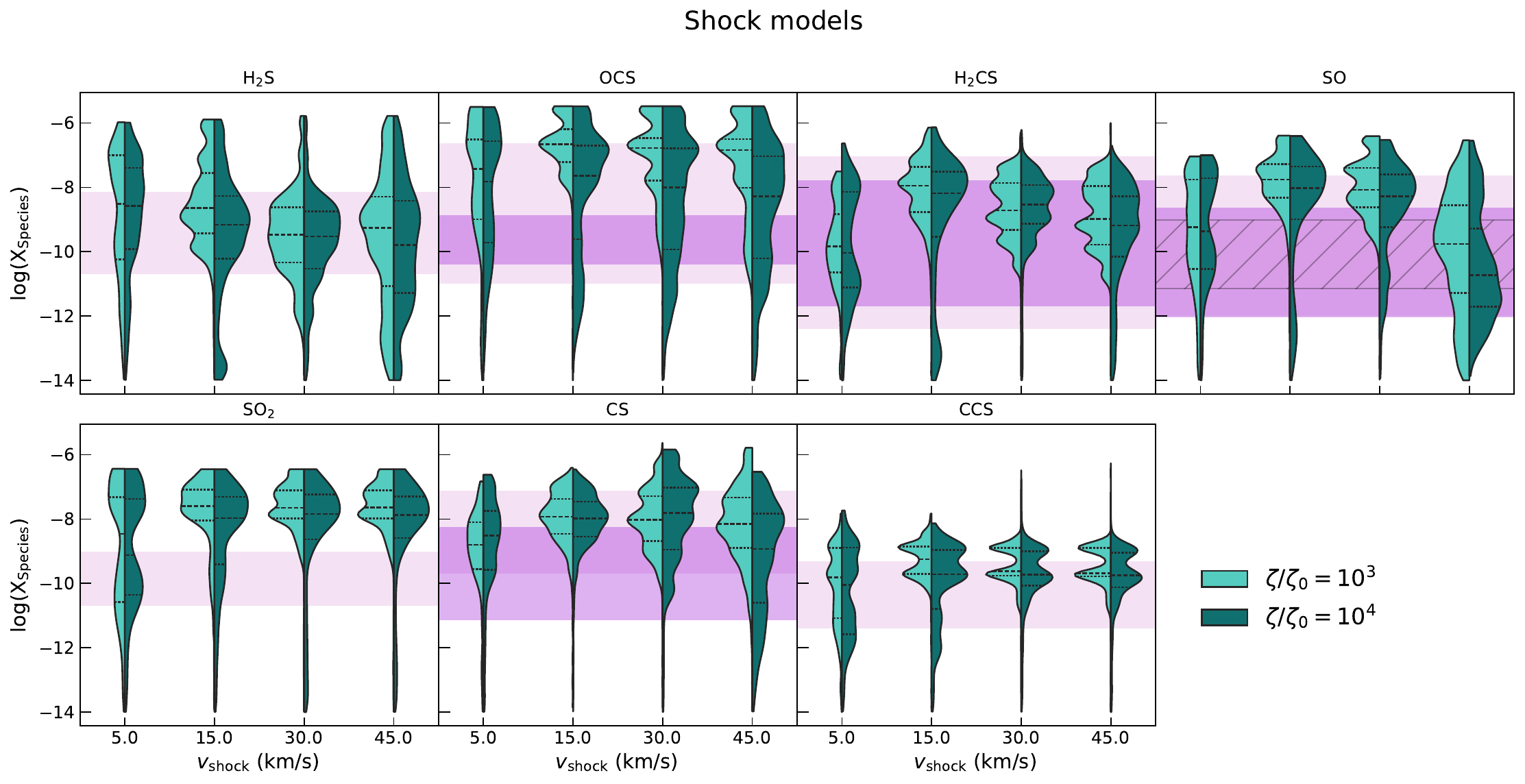}
    \caption{Violin plots of the fractional abundances in shock models as a function of the shock velocity (x-axis) and $\zeta/\zeta_0$ (shaded colours). The interquartiles of the data are shown. The observed abundances are shown as horizontal shaded purple areas. For species presenting a second component, the shaded area is darker. In the SO case (see text), we distinguish SO$_{\mathrm{comp1a}}$ (light purple) from SO$_{\mathrm{comp1b}}$ (dashed area).}
    \label{fig:shock_vel}
\end{figure}

\begin{figure}
    \centering
    \includegraphics[width=1\linewidth]{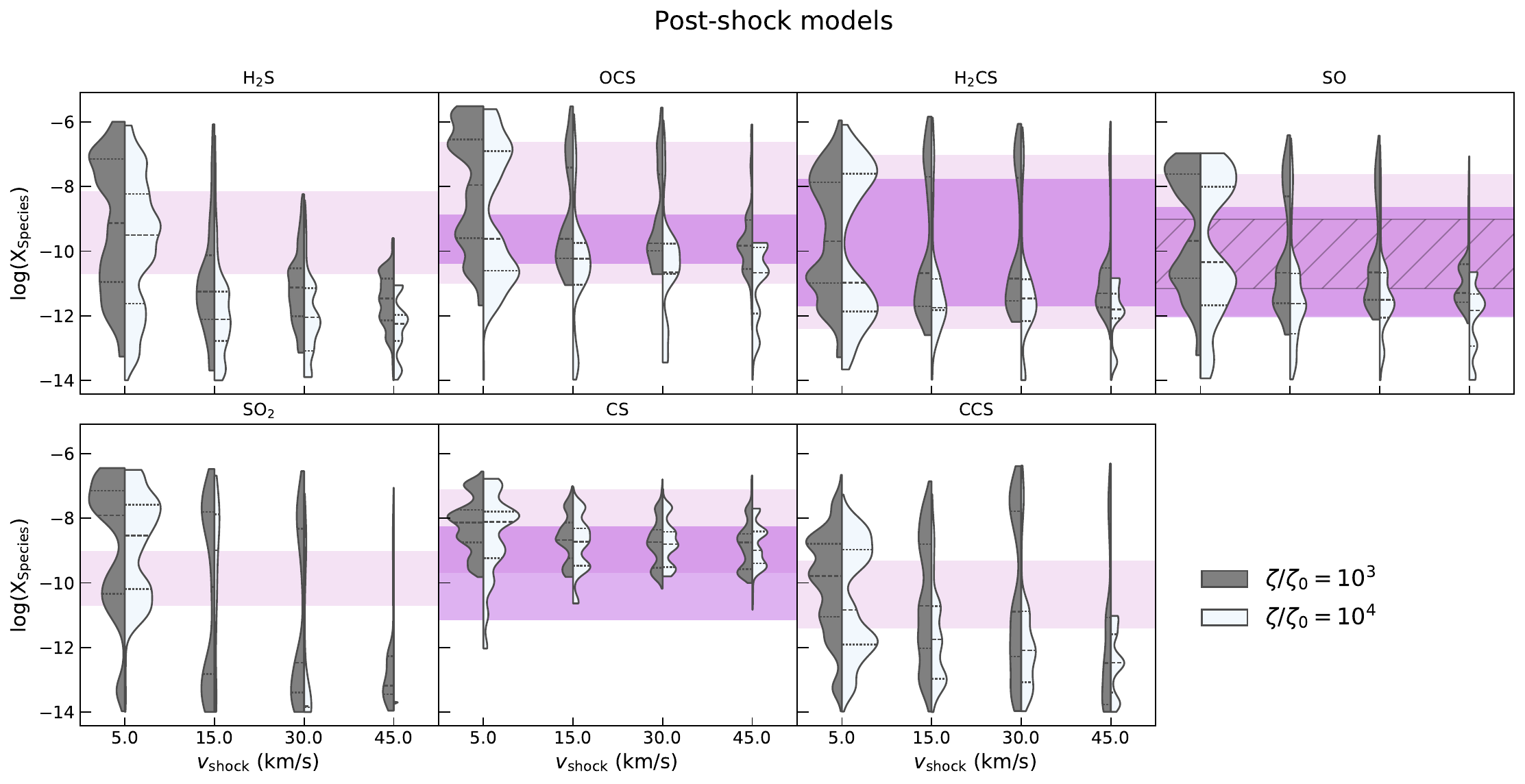}
    \caption{Same as Fig.~\ref{fig:shock_vel} but for post-shock models.}
    \label{fig:postshock_vel}
\end{figure}

Figure~\ref{fig:shock_vel} shows the modelled abundances as a function of the shock velocity.
The variation of the shock velocity from 5 \kms to 45 \kms does not critically affect the maximum abundances of sulphur-bearing species in shock models. We observe a slight increase for CS and CCS for the highest velocities, $v_{\mathrm{vel}}=30-45$ \kms, for \ce{H2CS} at $v_{\mathrm{vel}}=15$ \kms and for SO at the intermediate velocities, $v_{\mathrm{vel}}=15-30$ \kms.  However, the shock velocity does impact the distribution of the data, and hence the agreement with the observations. 
Similarly, Figure~\ref{fig:postshock_vel} shows the effect of the shock velocity on the abundances of the sulphur-bearing species in the post-shock models. Contrary to the shock models, the maximum abundance of \ce{H2S} decreases with an increasing $v_{\mathrm{vel}}$. This is however the only species affected. The shape of the distribution also drastically changes for any shock velocity higher than 5 \kms.

\paragraph{Pre-shock density ($n_{\mathrm{H,pre-shock}}$):}

\begin{figure}
    \centering
    \includegraphics[width=1\linewidth]{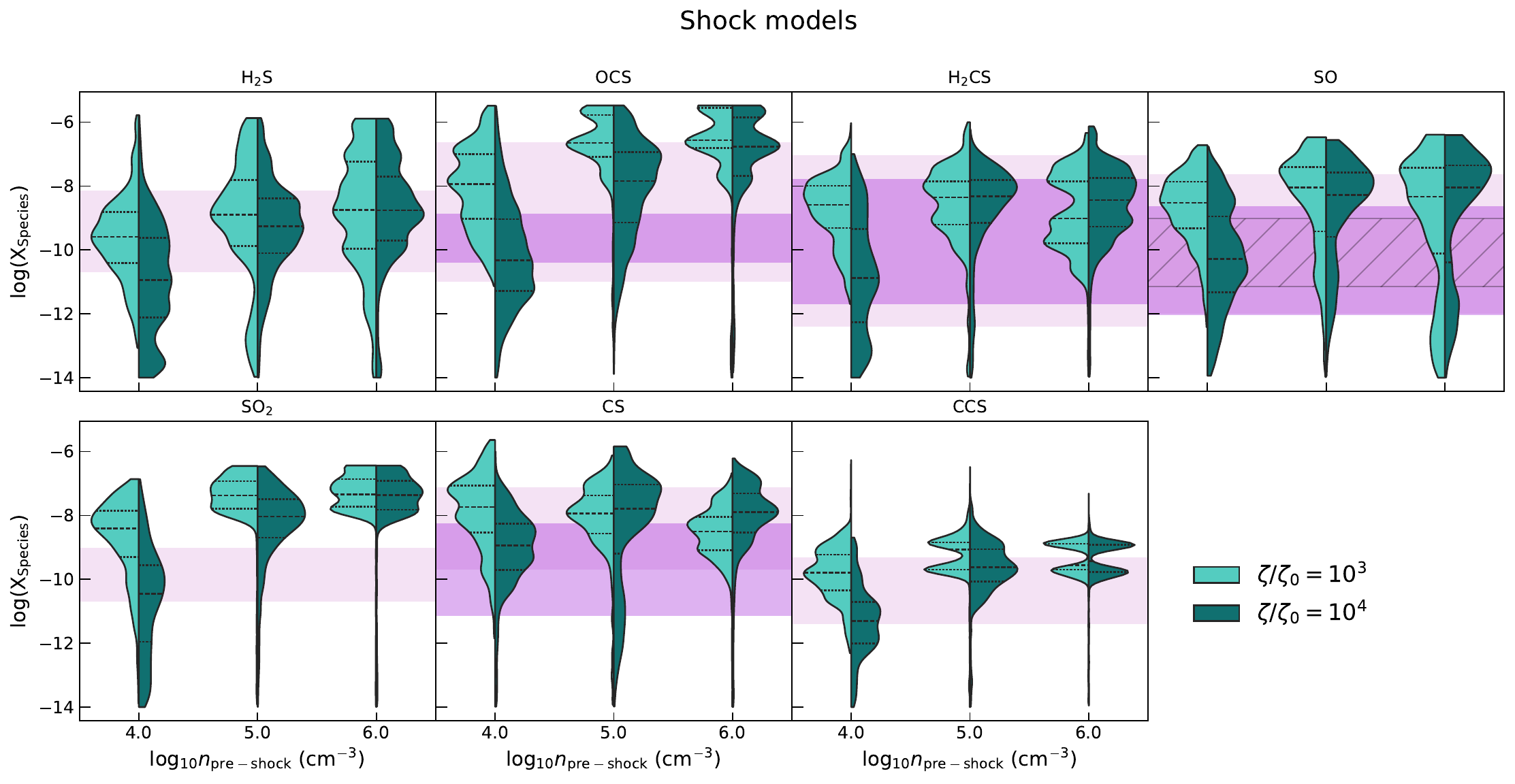}
    \caption{Violin plots of the fractional abundances in shock models as a function of the pre-shock density (x-axis) and $\zeta/\zeta_0$ (shaded colours). The interquartiles of the data are shown. The observed abundances are shown as horizontal shaded purple areas. For species presenting a second component, the purple shaded area is darker. In the SO case (see text), we distinguish SO$_{\mathrm{comp1a}}$ (light purple) from SO$_{\mathrm{comp1b}}$ (dashed area).}
    \label{fig:shock_dens}
\end{figure}

\begin{figure}
    \centering
    \includegraphics[width=1\linewidth]{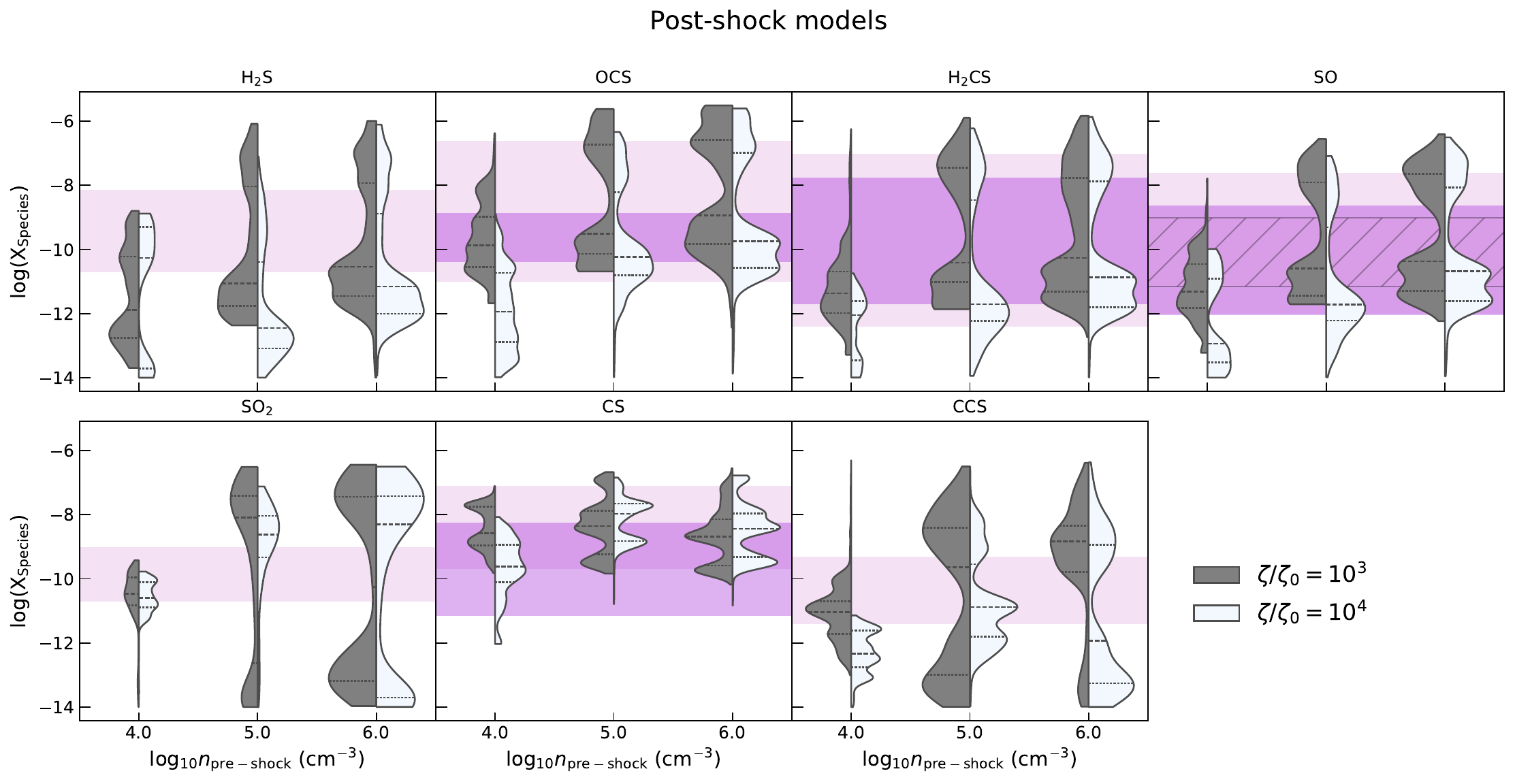}
    \caption{Same as Fig.~\ref{fig:shock_dens} but for post-shock models.}
    \label{fig:postshock_dens}
\end{figure}

Figures~\ref{fig:shock_dens} and ~\ref{fig:postshock_dens} show the modelled abundances of the sulphur-bearing species as a function of the pre-shock density, $n_{\mathrm{pre-shock}}$. In the shock models, as for the shock velocity, the maximum abundances are not much affected. We observe a slight increase with the increasing pre-shock density for OCS and \ce{SO2}, and a slight decrease in the case of CS and CCS at $n_{\mathrm{pre-shock}}=10^6$ \pcmc. The shape of the distribution is however impacted by a change in $n_{\mathrm{pre-shock}}$. 
In the post-shock models, the maximum abundances are the lowest for $n_{\mathrm{pre-shock}}=10^4$ \pcmc, for all the species, the most drastic changes being for \ce{H2S}, SO, and \ce{SO2}, with a change in up to three orders of magnitude. 

\end{appendix}

\end{document}